\documentstyle[12pt,epsfig]{article}

\newcommand{\tr}{\mathop{\mathrm{tr}}\nolimits}

\newcommand{\agt}{\,\rlap{\lower 3.5 pt \hbox{$\mathchar \sim$}} \raise 1pt
 \hbox {$>$}\,}
\newcommand{\alt}{\,\rlap{\lower 3.5 pt \hbox{$\mathchar \sim$}} \raise 1pt
 \hbox {$<$}\,}

\catcode`@=11
\newcount\@tempcntc
\def\@citex[#1]#2{\if@filesw\immediate\write\@auxout{\string\citation{#2}}\fi
  \@tempcnta\z@\@tempcntb\m@ne\def\@citea{}\@cite{\@for\@citeb:=#2\do
    {\@ifundefined
       {b@\@citeb}{\@citeo\@tempcntb\m@ne\@citea\def\@citea{,}{\bf ?}\@warning
       {Citation `\@citeb' on page \thepage \space undefined}}%
    {\setbox\z@\hbox{\global\@tempcntc0\csname b@\@citeb\endcsname\relax}%
     \ifnum\@tempcntc=\z@ \@citeo\@tempcntb\m@ne
       \@citea\def\@citea{,}\hbox{\csname b@\@citeb\endcsname}%
     \else
      \advance\@tempcntb\@ne
      \ifnum\@tempcntb=\@tempcntc
      \else\advance\@tempcntb\m@ne\@citeo
      \@tempcnta\@tempcntc\@tempcntb\@tempcntc\fi\fi}}\@citeo}{#1}}
\def\@citeo{\ifnum\@tempcnta>\@tempcntb\else\@citea\def\@citea{,}%
  \ifnum\@tempcnta=\@tempcntb\the\@tempcnta\else
   {\advance\@tempcnta\@ne\ifnum\@tempcnta=\@tempcntb \else \def\@citea{--}\fi
    \advance\@tempcnta\m@ne\the\@tempcnta\@citea\the\@tempcntb}\fi\fi}
\catcode`@=12

\begin{document}

\title{
\vskip-3cm{\baselineskip14pt
\centerline{\normalsize DESY 02-009\hfill ISSN 0418-9833}
\centerline{\normalsize hep-ph/0201234\hfill}
\centerline{\normalsize January 2002\hfill}
}
\vskip1.5cm
$J/\psi$ Inclusive Production in $\nu N$ Neutral-Current Deep-Inelastic
Scattering}
\author{
{\sc Bernd A. Kniehl and Lennart Zwirner}\\
{\normalsize II. Institut f\"ur Theoretische Physik, Universit\"at Hamburg,}\\
{\normalsize Luruper Chaussee 149, 22761 Hamburg, Germany}}

\date{}

\maketitle

\thispagestyle{empty}

\begin{abstract}
We calculate the cross section of $J/\psi$ inclusive production in
neutrino-nucleon deep-inelastic scattering via the weak neutral current within
the factorization formalism of nonrelativistic quantum chromodynamics.
Besides $J/\psi$ single production via the $Z$-gluon fusion mechanism, we also
consider $J/\psi$ plus hadron-jet associated production.
We take into account both direct production and feed-down from
directly-produced heavier charmonia.
We present theoretical predictions for the $J/\psi$ transverse-momentum and
rapidity distributions, which can be measured in the CHORUS and NOMAD
experiments at CERN, including conservative error estimates.
In order to interpret a recent CHORUS measurement of the total cross section,
we also estimate the contribution due to $J/\psi$ prompt production via
diffractive processes using the vector-meson dominance model.

\medskip

\noindent
PACS numbers: 12.38.-t, 13.60.Hb, 13.60.Le, 14.40.Gx
\end{abstract}

\newpage

\section{Introduction}
\label{sec:one}

Since its discovery in 1974, the $J/\psi$ meson has provided a useful
laboratory for quantitative tests of quantum chromodynamics (QCD) and, in
particular, of the interplay of perturbative and nonperturbative phenomena.
The factorization formalism of nonrelativistic QCD (NRQCD) \cite{bbl} provides
a rigorous theoretical framework for the description of heavy-quarkonium
production and decay.
This formalism implies a separation of short-distance coefficients, which can 
be calculated perturbatively as expansions in the strong-coupling constant
$\alpha_s$, from long-distance matrix elements (MEs), which must be extracted
from experiment.
The relative importance of the latter can be estimated by means of velocity
scaling rules, i.e.\ the MEs are predicted to scale with a definite power of
the heavy-quark ($Q$) velocity $v$ in the limit $v\ll1$.
In this way, the theoretical predictions are organized as double expansions in
$\alpha_s$ and $v$.
A crucial feature of this formalism is that it takes into account the complete
structure of the $Q\overline{Q}$ Fock space, which is spanned by the states
$n={}^{2S+1}L_J^{(c)}$ with definite spin $S$, orbital angular momentum $L$,
total angular momentum $J$, and colour multiplicity $c=1,8$.
In particular, this formalism predicts the existence of colour-octet (CO)
processes in nature.
This means that $Q\overline{Q}$ pairs are produced at short distances in
CO states and subsequently evolve into physical, colour-singlet (CS) quarkonia
by the nonperturbative emission of soft gluons.
In the limit $v\to0$, the traditional CS model (CSM) \cite{ber} is recovered.
The greatest triumph of this formalism was that it was able to correctly 
describe \cite{ebr} the cross section of inclusive charmonium
hadroproduction measured in $p\overline{p}$ collisions at the Fermilab
Tevatron \cite{abe}, which had turned out to be more than one order of
magnitude in excess of the theoretical prediction based on the CSM.

In order to convincingly establish the phenomenological significance of the
CO processes, it is indispensable to identify them in other kinds of
high-energy experiments as well.
Studies of charmonium production in $ep$ photoproduction, $ep$ and $\nu N$
deep-inelastic scattering (DIS), $e^+e^-$ annihilation, $\gamma\gamma$
collisions, and $b$-hadron decays may be found in the literature; see
Ref.~\cite{bra} and references cited therein.
Here, $N$ denotes a nucleon.
Furthermore, the polarization of charmonium, which also provides a sensitive
probe of CO processes, was investigated \cite{ben,bkl}.
Until very recently, none of these studies was able to prove or disprove the
NRQCD factorization hypothesis \cite{bbl}.
However, preliminary data of $\gamma\gamma\to J/\psi+X$ taken by the DELPHI
Collaboration \cite{delphi} at LEP2 provide first independent evidence for it
\cite{gg}.

In this paper, we revisit $J/\psi$ inclusive production in $\nu N$ DIS.
In particular, we consider the process $\nu+N\to\nu+J/\psi+X$, which is 
mediated via the weak neutral current (NC).
First experimental evidence for this process was delivered two decades ago by
the CERN-Dortmund-Heidelberg-Saclay (CDHS) Collaboration, who exposed an iron
target in the wide-band neutrino beam produced by protons from the CERN Super 
Proton Synchrotron (SPS) and identified the $J/\psi$ mesons through their
decays to $\mu^+\mu^-$ pairs \cite{cdhs}.
Similar experiments are being performed by the NuTeV Collaboration at Femilab
(Experiment E815) \cite{nutev} and by the CHORUS \cite{chorus} and NOMAD
\cite{nomad} Collaborations at CERN, using iron, lead, and iron targets,
respectively.
While NuTeV does not yet see evidence for NC production of $J/\psi$ mesons,
CHORUS nicely confirms the CDHS measurement.
On the other hand, NOMAD has not yet released any results on this.

The process $\nu+N\to\nu+J/\psi+X$ can either proceed diffractively or 
nondiffractively.
In diffractive processes, the target nucleons interact with the rest of the
process via exchanges of colourless objects with small spacelike virtualities
$t$, and they typically stay intact (elastic scattering) or turn into a
hadronic system of small invariant mass (diffractive dissociation).
On the other hand, in nondiffractive processes, they participate in the hard
scattering via their quark and gluon content and are thus destroyed
(deep-inelastic scattering).
In the case of elastic scattering off nucleons bound in compound nuclei, the
recoiling nuclei can emerge unharmed (coherent scattering) or break up
(quasielastic scattering).

The diffractive process $\nu+N\to\nu+J/\psi+N^\prime$, where $N^\prime$ 
denotes a system with little or no excitation, was studied in the framework of
the vector-meson dominance model (VDM) \cite{gai,kue,pol}.
The nondiffractive process $\nu+N\to\nu+J/\psi+X$ was treated in the parton
model of QCD endowed with two alternative approaches:
on the basis of the so-called $Z$-gluon fusion mechanism \cite{kue,pol,bar} in
the colour evaporation model based on local parton-hadron duality \cite{cem};
and in the CSM \cite{bai}.
For reviews, see Refs.~\cite{pol,seh}.

In all these processes, the $J/\psi$ mesons can be produced either directly or
via radiative or hadronic decays of heavier charmonia, such as $\chi_{cJ}$ and
$\psi^\prime$ mesons, where $J=0,1,2$.
The respective decay branching fractions are
$B(\chi_{c0}\to J/\psi+\gamma)=(0.66\pm0.18)\%$,
$B(\chi_{c1}\to J/\psi+\gamma)=(27.3\pm1.6)\%$,
$B(\chi_{c2}\to J/\psi+\gamma)=(13.5\pm1.1)\%$, and
$B(\psi^\prime\to J/\psi+X)=(55\pm5)\%$ \cite{pdg}.
It has become customary to collectively denote the $J/\psi$ mesons produced
directly or via the feed-down from heavier charmonia as prompt.
If the $\nu N$ centre-of-mass (CM) energy $\sqrt S$ is sufficiently large,
$\sqrt S>2M_B\approx10.6$~GeV, then $J/\psi$ mesons can also originate from 
weak decays of $b$ hadrons produced through the process
$\nu+N\to\nu+B+\bar B+X$ in diffractive or nondiffractive DIS.
The corresponding decay branching fraction is
$B(B\to J/\psi+X)=(1.16\pm0.10)\%$ \cite{pdg}.
In fixed-target experiments, the requisite incident-neutrino energy $E$ is
$E>(4M_B^2-m^2)/(2m)\approx59$~GeV, where $m$ is the nucleon mass.
In the case of the CERN wide-band neutrino beam, only an insignificant
fraction, namely 7\%, of the total flux satisfies this criterion \cite{nomad}.
Therefore and because of the smallness of $B(B\to J/\psi+X)$, $J/\psi$ 
production through $b$-hadron decay should be greatly suppressed in the CHORUS
and NOMAD experiments and is not considered here.

In this paper, we present complete and up-to-date predictions for prompt
$J/\psi$ production in $\nu N$ NC DIS, including both the diffractive and 
nondiffractive contributions, which can be readily compared with available and 
expected CHORUS and NOMAD data.
Specifically, we calculate the nondiffractive cross section of
$\nu+N\to\nu+J/\psi+X$, where $X$ denotes the nucleon remnant and possibly one
additional hadron jet, to lowest order (LO) in the framework of the NRQCD
factorization formalism \cite{bbl}, and we reanalyze the diffractive cross 
section of $\nu+N\to\nu+J/\psi+N^\prime$ in the VDM adopting the results of
Refs.~\cite{kue,pol} complemented with new input information.
As for the nondiffractive direct contribution, our analysis is, in a way,
complementary to the one of Ref.~\cite{ep}, where $e+p\to e+J/\psi+X$ via
photons with large virtualities $Q^2$ was investigated:
while the photon-quark coupling is purely vectorial, $Z$ bosons couple to
quarks mainly through axial-vector couplings, the vector-coupling strengths
being numerically small.
However, by charge-conjugation invariance, the leading CSM prediction only
involves vector couplings and may thus be obtained from Ref.~\cite{ep} by
appropriate substitutions \cite{bai}.
As mentioned above, the nondiffractive prompt CS contribution was already
studied in Ref.~\cite{bai}, two decades ago.
However, since the authors of that paper did not present analytic results and
used input parameters that are now obsolete, it is indispensable to repeat 
this analysis.

The leading MEs of the $S$-wave charmonia $\psi=J/\psi,\psi^\prime$ are
$\left\langle{\cal O}^\psi\left[{}^3\!S_1^{(1)}\right]\right\rangle$,
$\left\langle{\cal O}^\psi\left[{}^1\!S_0^{(8)}\right]\right\rangle$,
$\left\langle{\cal O}^\psi\left[{}^3\!S_1^{(8)}\right]\right\rangle$, and
$\left\langle{\cal O}^\psi\left[{}^3\!P_J^{(8)}\right]\right\rangle$, with
$J=0,1,2$, while those of the $P$-wave charmonia $\chi_{cJ}$ are
$\left\langle{\cal O}^{\chi_{cJ}}\left[{}^3\!P_J^{(1)}\right]\right\rangle$ 
and
$\left\langle{\cal O}^{\chi_{cJ}}\left[{}^3\!S_1^{(8)}\right]\right\rangle$.
They satisfy the multiplicity relations
\begin{eqnarray}
\left\langle{\cal O}^\psi\left[{}^3\!P_J^{(8)}\right]\right\rangle
&=&(2J+1)\left\langle{\cal O}^\psi\left[{}^3\!P_0^{(8)}\right]\right\rangle,
\nonumber\\
\left\langle{\cal O}^{\chi_{cJ}}\left[{}^3\!P_J^{(1)}\right]\right\rangle
&=&(2J+1)
\left\langle{\cal O}^{\chi_{c0}}\left[{}^3\!P_0^{(1)}\right]\right\rangle,
\nonumber\\
\left\langle{\cal O}^{\chi_{cJ}}\left[{}^3\!S_1^{(8)}\right]\right\rangle
&=&(2J+1)
\left\langle{\cal O}^{\chi_{c0}}\left[{}^3\!S_1^{(8)}\right]\right\rangle,
\label{eq:mul}
\end{eqnarray}
which follow to LO in $v$ from heavy-quark spin symmetry.

At ${\cal O}(\alpha^2\alpha_s)$, where $\alpha$ is Sommerfeld's fine-structure
constant, the $Z$-gluon fusion mechanism \cite{kue,pol,bar} is realized by the
CO partonic subprocesses $\nu+g\to\nu+c\overline{c}[n]$, where
$n={}^1\!S_0^{(8)},{}^3\!S_1^{(8)},{}^3\!P_J^{(8)}$ \cite{tor}.
In order to enable the production of the CS Fock states
$n={}^3\!S_1^{(1)},{}^3\!P_J^{(1)}$, we need to allow for one additional gluon
in the final state.
The resulting LO CS partonic subprocesses, $\nu+g\to\nu+c\overline{c}[n]+g$,
are of ${\cal O}(\alpha^2\alpha_s^2)$ (see Fig.~\ref{fig:fey}) \cite{bai}.

The cross sections of
$\nu+g\to\nu+c\overline{c}\left[{}^3\!P_J^{(1)}\right]+g$ suffer from infrared
(IR) singularities in the limit of the outgoing gluon being soft.
This reflects a well-known conceptual deficiency of the CSM \cite{ber}.
In the NRQCD framework \cite{bbl}, these IR singularities are factorized at
some cut-off scale and absorbed into redefinitions of the CO MEs
$\left\langle{\cal O}^{\chi_{cJ}}\left[{}^3\!S_1^{(8)}\right]\right\rangle$
that multiply the cross section of
$\nu+g\to\nu+c\overline{c}\left[{}^3\!S_1^{(8)}\right]$.
In turn, these CO MEs become scale dependent at ${\cal O}(\alpha_s)$ and
satisfy appropriate evolution equations.
The factorization is conveniently performed in dimensional regularization
\cite{che}.
In Ref.~\cite{bai}, these IR singularities were not encountered because the
hadronic recoil energy $E_h$ was required to exceed some minimum value
$E_h^{\rm min}$.
However, a rapid rise of cross section with decreasing value of
$E_h^{\rm min}$ was observed.

In the same order, we also have the CO partonic subprocesses
$\nu+a\to\nu+c\overline{c}[n]+a$, where $a=q,\overline{q},g$, with $q=u,d,s$,
and $n={}^1\!S_0^{(8)},{}^3\!S_1^{(8)},{}^3\!P_J^{(8)}$ (see
Fig.~\ref{fig:fey}).
The cross sections of the latter are plagued by collinear singularities in the
limit of vanishing $c\overline{c}$ transverse momentum $p_T^\star$ in the
$Z^\star N$ CM frame.
Here, $Z^\star$ denotes the virtual $Z$ boson.
These singularities could be avoided by introducing an appropriate cut on
$p_T^\star$.
In a full next-to-leading-order (NLO) analysis, they would be factorized at
mass scale $\mu_f$ and absorbed into the bare parton density functions (PDFs)
of the nucleon $N$ so as to renormalize the latter.

In this paper, we provide analytical results for the cross sections of all the
$2\to2$ and $2\to3$ partonic subprocesses enumerated above to LO within the
NRQCD factorization formalism \cite{bbl}.
In the $2\to2$ case, we found agreement with Eqs.~(8) and (9) of 
Ref.~\cite{tor}\footnote{%
There is a typographical error in the last equation of Eq.~(9):
$\left(Q^2+4m_c^2\right)^2$ in the denominator should be replaced by
$\left(Q^2+4m_c^2\right)^4$.}, so that there is no need to list our formulas
for the squared transition-matrix elements.
We then calculate the inclusive cross section of prompt $J/\psi$ production in
nondiffractive $\nu N$ NC DIS under CHORUS and NOMAD kinematic conditions to
LO in NRQCD and the CSM.
This involves the $2\to2$ CO processes of $J/\psi$, $\chi_{cJ}$, and
$\psi^\prime$ production and the $2\to3$ CS processes of $J/\psi$ and
$\psi^\prime$ production.
As explained above, the $2\to3$ CS processes of $\chi_{cJ}$ production lie
beyond the scope of the CSM.
We also consider the distributions in the $J/\psi$ transverse momentum $p_T$ 
and rapidity $y$ in the laboratory frame, which should be experimentally 
accessible.

\begin{table}[ht]
\begin{center}
\caption{Relative importance, measured in powers of $\alpha_s$ and
$v^2\approx\alpha_s$, of the various nondiffractive $J/\psi$ production
channels according to the NRQCD counting rules.
The entries are normalized to the direct $2\to3$ CS contribution, which is of
${\cal O}\left(\alpha^2\alpha_s^2v^3\right)$.}
\label{tab:one}
\medskip
\begin{tabular}{|c|cc|}
\hline\hline
& $J/\psi$, $\psi^\prime$ & $\chi_{cJ}$ \\
\hline
$2\to2$ CO & $v^4/\alpha_s\approx\alpha_s$ & $v^2/\alpha_s\approx1$ \\
$2\to3$ CS & 1 & $v^2\approx\alpha_s$ \\
$2\to3$ CO & $v^4\approx\alpha_s^2$ & $v^2\approx\alpha_s$ \\
\hline\hline
\end{tabular}
\end{center}
\end{table}
The relative importance, measured in powers of $\alpha_s$ and
$v^2\approx\alpha_s$, of the various nondiffractive $J/\psi$ production
channels according to the NRQCD counting rules is represented in 
Table~\ref{tab:one}.
The entries are normalized to the direct $2\to3$ CS contribution, which is of
${\cal O}\left(\alpha^2\alpha_s^2v^3\right)$.
In addition, the feed-down channels are suppressed by the respective branching
fractions.
In fixed-target experiments, which are typically performed at low values of
$\sqrt S$ \cite{cdhs,nutev,chorus,nomad}, phase-space suppression comes in as
another limiting factor.
This affects the $2\to3$ processes more severely than the $2\to2$ ones.
In fact, as will become apparent in Section~\ref{sec:three}, this is the 
reason why, in the case of direct $J/\psi$ and $\psi^\prime$ production, the
$2\to2$ CO contributions greatly exceed the $2\to3$ CS ones, although they are
formally suppressed according to Table~\ref{tab:one}.
If events of nondiffractive direct $J/\psi$ and $\psi^\prime$ production could
be identified experimentally, this would lend itself to a powerful
discriminator between the CSM and NRQCD.
Obviously, the $2\to3$ CO contributions, which are not included in our
numerical analysis for the reasons explained above, would only provide minor
corrections to their $2\to2$ counterparts, which make up the bulk of the NRQCD
prediction.

This paper is organized as follows.
In Section~\ref{sec:two}, we present, in analytic form, the cross sections of
the partonic subprocesses enumerated above and explain how to calculate from
them the total cross section of prompt $J/\psi$ production in nondiffractive
$\nu N$ NC DIS as well as its $p_T$ and $y$ distributions.
Lengthy expressions are relegated to the Appendix.
For the reader's convenience, we also recall the cross-section formulas for
prompt $J/\psi$ production in diffractive $\nu N$ NC DIS. 
In Section~\ref{sec:three}, we present our numerical results and compare them
with recent CHORUS data.
Our conclusions are summarized in Section~\ref{sec:four}.

\section{Analytic results}
\label{sec:two}

In this section, we present our analytic results for the cross sections of
$\nu+N\to\nu+H+j+X$ in nondiffractive DIS, which proceeds through the
$2\to3$ partonic subprocesses $\nu+a\to\nu+c\overline{c}[n]+a$ mentioned in
Section~\ref{sec:one}.
Here, $H=J/\psi,\chi_{cJ},\psi^\prime$, the hadron jet $j$ arises from the
fragmentation of the additional final-state parton $a$, and $X$ is the nucleon
remnant.
In our numerical analysis, we only consider the CS processes of $J/\psi$ and
$\psi^\prime$ production and integrate their cross sections over all
kinematically allowed values of the $j$ three-momentum.
The CS processes of $\chi_{cJ}$ production and the CO processes are provided
for future applications.
At the end of this section, we also present formulas for the cross sections of
the processes $\nu+N\to\nu+H+X$ and $\nu+N\to\nu+H+N^\prime$.

We work at LO in the parton model of QCD with $n_f=3$ active quark flavours
and employ the NRQCD factorization formalism \cite{bbl} to describe the
formation of the $H$ meson.
We start by defining the kinematics.
As indicated in Fig.~\ref{fig:kin}, we denote the four-momenta of the incoming
neutrino and nucleon and the outgoing neutrino, $H$ meson, and hadron jet
by $k$, $P$, $k^\prime$, $p_H$, and $p^\prime$, respectively.
The parton struck by the virtual $Z$ boson carries four-momentum $p=xP$.
We neglect the masses of the nucleon and the light quarks, call the one
of the $H$ meson $M$, and take the charm-quark mass to be $m_c=M/2$.
In our approximation, the nucleon remnant $X$ has zero invariant mass,
$M_X^2=(P-p)^2=0$.
The CM energy square of the $\nu N$ collision is $S=(k+P)^2$.
The virtual $Z$ boson has four-momentum $q=k-k^\prime$.
As usual, we define $Q^2=-q^2>0$, $y=q\cdot P/k\cdot P$, and the inelasticity 
variable $z=p_H\cdot P/q\cdot P$.
In the nucleon rest frame, $y$ and $z$ measure the relative neutrino energy
loss and the fraction of the $Z^\star$ energy transferred to the $H$ meson,
respectively.
The $Z^\star N$ CM energy square is $W^2=(q+P)^2=yS-Q^2$.
The system $X^\prime$ consisting of $j$ and $X$ has invariant mass square
$M_{X^\prime}^2=(q+P-p_H)^2=(1-x)y(1-z)S$.
As usual, we define the partonic Mandelstam variables as
$\hat s=(q+p)^2=xyS-Q^2$, $\hat t=(q-p_H)^2=-xy(1-z)S$, and
$\hat u=(p-p_H)^2=M^2-xyzS$.
By four-momentum conservation, we have $\hat s+\hat t+\hat u=M^2-Q^2$.
In the $Z^\star N$ CM frame, the $H$ meson has transverse momentum and
rapidity
\begin{eqnarray}
p_T^\star&=&\frac{\sqrt{\hat t\left(\hat s\hat u+Q^2M^2\right)}}{\hat s+Q^2},
\label{eq:ptcms}\\
y_H^\star&=&\frac{1}{2}\ln\frac{\hat s\left(M^2-\hat u\right)}
{\hat s\left(M^2-\hat t\right)+Q^2M^2}+\frac{1}{2}\ln\frac{W^2}{\hat s},
\label{eq:ycms}
\end{eqnarray}
respectively.
Here and in the following, we denote the quantities referring to the
$Z^\star N$ CM frame by an asterisk.
The second term on the right-hand side of Eq.~(\ref{eq:ycms}) originates from 
the Lorentz boost from the $Z^\star a$ CM frame to the $Z^\star N$ one.
Here, $y_H^\star$ is taken to be positive in the direction of the
three-momentum of the virtual $Z$ boson.

In the parton model, the nucleon is characterized by its PDFs
$f_{a/N}(x,\mu_f)$, and, at LO, an outgoing parton may be identified with a
hadron jet.
Thus, we have
\begin{equation}
d\sigma(\nu+N\to\nu+H+j+X)
=\int_0^1dx\sum_af_{a/N}(x,\mu_f)d\sigma(\nu+a\to\nu+H+a),
\label{eq:par}
\end{equation}
where $a=u,\overline{u},d,\overline{d},s,\overline{s},g$.
Furthermore, according to the NRQCD factorization formalism \cite{bbl}, we
have
\begin{equation}
d\sigma(\nu+a\to\nu+H+a)=\sum_n\left\langle{\cal O}^H[n]\right\rangle
d\sigma(\nu+a\to\nu+c\overline{c}[n]+a),
\label{eq:fac}
\end{equation}
where, to LO in $v$,
$n={}^3\!S_1^{(1)},{}^1\!S_0^{(8)},{}^3\!S_1^{(8)},{}^3\!P_J^{(8)}$ for
$H=J/\psi,\psi^\prime$ and $n={}^3\!P_J^{(1)},{}^3\!S_1^{(8)}$ for
$H=\chi_{cJ}$.

Decomposing the transition-matrix element of the partonic subprocess
$\nu+a\to\nu+c\overline{c}[n]+a$ into a leptonic part,
\begin{equation}
{\cal T}^\mu(\nu\to\nu+Z^\star)=-i\frac{g}{q^2-M_Z^2}
\overline{u}(k^\prime)\gamma^\mu\frac{1-\gamma_5}{2}u(k),
\end{equation}
where $g=2^{1/4}G_F^{1/2}M_Z$, with $G_F$ being Fermi's constant, and a
hadronic one, ${\cal T}^\mu(Z^\star+a\to c\overline{c}[n]+a)$, from
which the virtual $Z$-boson leg is amputated, we can write its cross section
as
\begin{eqnarray}
d\sigma(\nu+a\to\nu+c\overline{c}[n]+a)
&=&\frac{1}{2xS}\,\frac{1}{2N_a}\,\frac{g^2}{\left(q^2-M_Z^2\right)^2}\, 
\tr\left(\not k\gamma^\nu\frac{1-\gamma_5}{2}\not k^\prime\gamma^\mu
\frac{1-\gamma_5}{2}\right)H_{\mu\nu}
\nonumber\\
&&{}\times d{\mathrm PS}_3(k+p;k^\prime,p_H,p^\prime),
\label{eq:dec}
\end{eqnarray}
where $N_q=N_{\overline{q}}=N_c=3$ and $N_g=(N_c^2-1)$ are the colour
multiplicities of the partons $a$ and the hadronic tensor $H^{\mu\nu}$ is
obtained by summing the absolute square of
${\cal T}^\mu(Z^\star+a\to c\overline{c}[n]+a)$ over the spin and colour
states of the incoming and outgoing partons $a$.
Here and in the following, we employ the Lorentz-invariant phase-space measure
\begin{equation}
d{\mathrm PS}_n(p;p_1,\ldots,p_n)
=(2\pi)^4\delta^{(4)}\left(p-\sum_{i=1}^np_i\right)\prod_{i=1}^n
\frac{d^3p_i}{(2\pi)^32p_i^0}.
\end{equation}
The first factor on the right-hand side of Eq.~(\ref{eq:dec}) stems from the
flux and the second one from the average over the spin and colour states of
the incoming parton.
Integrating over the azimuthal angle of the outgoing neutrino, we may simplify
Eq.~(\ref{eq:dec}) to become
\begin{equation}
d\sigma(\nu+a\to\nu+c\overline{c}[n]+a)
=\frac{1}{2xS}\,\frac{1}{2N_a}\left(\frac{g}{4\pi}\right)^2
L^{\mu\nu}H_{\mu\nu}\frac{dy}{y}\,\frac{Q^2dQ^2}{\left(Q^2+M_Z^2\right)^2}
d{\mathrm PS}_2(q+p;p_H,p^\prime),
\label{eq:red}
\end{equation}
where 
\begin{equation}
L^{\mu\nu}=\frac{1+(1-y)^2}{y}\epsilon_T^{\mu\nu}
-\frac{4(1-y)}{y}\epsilon_L^{\mu\nu}+(2-y)\epsilon_A^{\mu\nu},
\label{eq:lep}
\end{equation}
with
\begin{eqnarray}
\epsilon_T^{\mu\nu}&=&-g^{\mu\nu}+\frac{1}{q\cdot p}(q^\mu p^\nu+p^\mu q^\nu)
-\frac{q^2}{(q\cdot p)^2}p^\mu p^\nu,
\nonumber\\
\epsilon_L^{\mu\nu}&=&\frac{1}{q^2}\left(q-\frac{q^2}{q\cdot p}p\right)^\mu
\left(q-\frac{q^2}{q\cdot p}p\right)^\nu,
\nonumber\\
\epsilon_A^{\mu\nu}&=&\frac{i}{q\cdot p}\epsilon^{\mu\nu\rho\sigma}
q_{\rho}p_{\sigma}
\end{eqnarray}
is the leptonic tensor.
Here, we adopt the convention $\epsilon_{0123}=1$.
The cross section of $\overline{\nu}+a\to\overline{\nu}+c\overline{c}[n]+a$
emerges from Eq.~(\ref{eq:red}) through crossing symmetry, by flipping the 
sign of the last term on the right-hand side of Eq.~(\ref{eq:lep}).
In the following, the symbol $\nu$ collectively denotes neutrinos and
antineutrinos.

We evaluate the cross sections of the relevant partonic subprocesses
$\nu+a\to\nu+c\overline{c}[n]+a$ from Eq.~(\ref{eq:red}) applying the
covariant-projector method of Ref.~\cite{pet}.
Our results can be written in the form
\begin{eqnarray}
\frac{d^3\sigma}{dy\,dQ^2\,d\hat t}
(\nu+a\to\nu+c\overline{c}[n]+a)
&=&F_a[n]\left[\frac{1+(1-y)^2}{y}T_a[n]-\frac{4(1-y)}{y}L_a[n]\right.
\nonumber\\
&&{}+
\left.\vphantom{\frac{1+(1-y)^2}{y}}
(2-y)A_a[n]\right],
\label{eq:res}
\end{eqnarray}
where $F_a[n]$, $T_a[n]$, $L_a[n]$, and $A_a[n]$ are functions of $\hat s$,
$\hat t$, $\hat u$, and $Q^2$, which are listed in the Appendix.
We combined the results proportional to the CO MEs
$\left\langle{\cal O}^\psi\left[{}^3\!P_J^{(8)}\right]\right\rangle$ and
$\left\langle{\cal O}^{\chi_{cJ}}\left[{}^3\!S_1^{(8)}\right]\right\rangle$
exploiting the multiplicity relations of Eq.~({\ref{eq:mul}).
The $T_a[n]$ and $L_a[n]$ functions involve terms proportional to $v_iv_j$ or
$a_ia_j$, while the $A_a[n]$ functions involve terms proportional to $v_ia_j$,
where $i,j=q,c$.
Here, $v_q=I_q^3-2e_q\sin^2\theta_w$ and $a_q=I_q^3$ are the $Zq\overline{q}$
vector and axial-vector couplings, respectively, where $I_q^3$ is the third
component of weak isospin of the left-handed component of quark $q$, $e_q$ is
the fractional electric charge of the latter, and $\theta_w$ is the weak
mixing angle.
We recover our result for the cross section of $e+a\to e+J/\psi+a$, given
in Eq.~(13) of Ref.~\cite{ep}, by substituting in Eq.~(\ref{eq:res}) $g=e$,
$v_q=e_q$, $v_c=e_c$, $a_q=a_c=0$, and $M_Z=0$, where $e=\sqrt{4\pi\alpha}$ is
the electron-charge magnitude.

For the CO Fock states $n={}^1\!S_0^{(8)},{}^3\!S_1^{(8)},{}^3\!P_J^{(8)}$,
the cross sections of Eq.~(\ref{eq:res}) exhibit collinear singularities in
the limit $\hat t\to0$.
According to the factorization theorem, the limiting expressions must coincide
with the respective $\nu+g\to\nu+c\overline{c}[n]$ cross sections \cite{tor}
multiplied by the spacelike $a\to g$ splitting functions.
This provides another nontrivial check for our results.

It is interesting to observe that the cross sections of
$\nu+g\to\nu+c\overline{c}\left[{}^3\!P_J^{(1)}\right]+g$ vanish in the limit
$y\to0$ and $Q^2\to0$.

Inserting Eq.~(\ref{eq:fac}) in Eq.~(\ref{eq:par}) and including the maximum
boundaries of the integrations over $x$ and $\hat t$, we obtain
\begin{eqnarray}
\lefteqn{\frac{d^2\sigma}{dy\,dQ^2}(\nu+N\to\nu+H+j+X)
=\int_{(Q^2+M^2)/(yS)}^1dx
\int_{-(\hat s+Q^2)(\hat s-M^2)/\hat s}^0d\hat t}
\nonumber\\
&&{}\times
\sum_af_{a/N}(x,\mu_f)\sum_n\left\langle{\cal O}^H[n]\right\rangle
\frac{d^3\sigma}{dy\,dQ^2\,d\hat t}(\nu+a\to\nu+c\overline{c}[n]+a),
\label{eq:dif}
\end{eqnarray}
where
$\left(d^3\sigma/dy\,dQ^2\,d\hat t\right)(\nu+a\to\nu+c\overline{c}[n]+a)$
is given by Eq.~(\ref{eq:res}).
The kinematically allowed ranges of $S$, $y$, and $Q^2$ are $S>M^2$,
$M^2/S<y<1$, and $0<Q^2<yS-M^2$, respectively.
We then evaluate the cross section of prompt $J/\psi$ production as
\begin{eqnarray}
\lefteqn{d\sigma(\nu+N\to\nu+(J/\psi)_{\rm prompt}+j+X)
=d\sigma(\nu+N\to\nu+J/\psi+j+X)}
\nonumber\\
&&{}+\sum_{J=0}^2B(\chi_{cJ}\to J/\psi+\gamma)
d\sigma(\nu+N\to\nu+\chi_{cJ}+j+X)
\nonumber\\
&&{}+B(\psi^\prime\to J/\psi+X)d\sigma(\nu+N\to\nu+\psi^\prime+j+X).
\end{eqnarray}

The distributions in $y$ and $Q^2$ can be evaluated from Eq.~(\ref{eq:dif}) as
it stands.
It is also straightforward to obtain the distributions in $p_T^\star$ and
$y_H^\star$, given in Eqs.~(\ref{eq:ptcms}) and (\ref{eq:ycms}),
respectively, by accordingly redefining and reordering the integration
variables in Eq.~(\ref{eq:dif}).
The distribution in the $J/\psi$ azimuthal angle $\phi^\star$ in the
$Z^\star N$ CM frame is constant.

The evaluation of the distributions in the $J/\psi$ transverse momentum $p_T$,
rapidity $y_H$, and azimuthal angle $\phi$ in the laboratory frame is somewhat
more involved.
Choosing a suitable coordinate system in the $Z^\star N$ CM frame, we have
\begin{eqnarray}
(k^\star)^\mu&=&\frac{S-Q^2}{2W}\left(
\begin{array}{c}
1\\
\sin\psi^\star\\
0\\
\cos\psi^\star
\end{array}
\right),
\qquad
(k^{\prime\star})^\mu=\frac{S-W^2}{2W}\left(
\begin{array}{c}
1\\
\sin\theta^\star\\
0\\
\cos\theta^\star
\end{array}
\right),
\nonumber\\
(q^\star)^\mu&=&\frac{1}{2W}\left(
\begin{array}{c}
W^2-Q^2\\
0\\
0\\
W^2+Q^2
\end{array}
\right),
\qquad
(P^\star)^\mu=\frac{W^2+Q^2}{2W}\left(
\begin{array}{c}
1\\
0\\
0\\
-1
\end{array}
\right),
\nonumber\\
(p_H^\star)^\mu&=&\left(
\begin{array}{c}
m_T^\star\cosh y_H^\star\\
p_T^\star\cos\phi^\star\\
p_T^\star\sin\phi^\star\\
m_T^\star\sinh y_H^\star
\end{array}
\right),
\end{eqnarray}
where\footnote{The formula for $\cos\theta^\star$ given in Ref.~\cite{ep}
contains a sign error, which was absent in the preprint version of that
paper.}
\begin{eqnarray}
\cos\psi^\star&=&\frac{2SW^2}{(S-Q^2)(W^2+Q^2)}-1,\qquad
\cos\theta^\star=1-\frac{2SQ^2}{(S-W^2)(W^2+Q^2)},
\nonumber\\
m_T^\star&=&\sqrt{M^2+\left(p_T^\star\right)^2}.
\end{eqnarray}
On the other hand, in the laboratory frame, which coincides with the nucleon
rest frame, we have
\begin{eqnarray}
k^\mu&=&E\left(
\begin{array}{c}
1\\
0\\
0\\
1
\end{array}
\right),
\qquad
(k^\prime)^\mu=E^\prime\left(
\begin{array}{c}
1\\
\sin\theta\\
0\\
\cos\theta
\end{array}
\right),
\nonumber\\
q^\mu&=&\left(
\begin{array}{c}
q^0\\
-q\sin\psi\\
0\\
q\cos\psi
\end{array}
\right),
\qquad
P^\mu=m\left(
\begin{array}{c}
1\\
0\\
0\\
0
\end{array}
\right),
\nonumber\\
p_H^\mu&=&\left(
\begin{array}{c}
m_T\cosh y_H\\
p_T\cos\phi\\
p_T\sin\phi\\
m_T\sinh y_H
\end{array}
\right),
\label{eq:rest}
\end{eqnarray}
where
\begin{eqnarray}
E&=&\frac{S-m^2}{2m},\qquad
E^\prime=\frac{S-W^2-Q^2}{2m},\qquad
\cos\theta=1-\frac{2Q^2m^2}{(S-m^2)(S-W^2-Q^2)},
\nonumber\\
q^0&=&\frac{W^2+Q^2-m^2}{2m},\qquad
q=\sqrt{Q^2+(q^0)^2},
\nonumber\\
\cos\psi&=&\frac{1}{q}\left(\frac{W^2+Q^2-m^2}{2m}
+\frac{Q^2m}{S-m^2}\right),\qquad
m_T=\sqrt{M^2+p_T^2}.
\label{eq:res0}
\end{eqnarray}
Notice that $y_H$ is taken to be positive in the direction of the
three-momentum of the incoming neutrino.
Without loss of generality, we may require that $0\le\psi^\star,\psi\le\pi$,
for, otherwise, we can achieve this by rotating the respective coordinate
systems by $180^\circ$ around the $z$ axis.
We can then evaluate $p_T$, $y_H$, and $\phi$ from $p_T^\star$, $y_H^\star$,
and $\phi^\star$ as
\begin{eqnarray}
p_T&=&\sqrt{\left(p_T^\star\right)^2+A\left(A-2p_T^\star\cos\phi^\star\right)},
\label{eq:ptlab}
\\
y_H&=&y_H^\star+\ln\frac{(W^2+Q^2)m_T^\star}{\sqrt SWm_T}
+\frac{1}{2}\ln\frac{S}{m^2},
\label{eq:ylab}
\\
\cos\phi&=&\frac{p_T^\star\cos\phi^\star-A}{p_T},
\label{eq:philab}
\end{eqnarray}
where 
\begin{equation}
A=\frac{m_T^\star\exp(y_H^\star)\sin\psi^\star}{1+\cos\psi^\star}
=\sqrt{\frac{Q^2(S-W^2-Q^2)}{SW^2}}m_T^\star\exp(y_H^\star).
\label{eq:a}
\end{equation}
The third term on the right-hand side of Eq.~(\ref{eq:ylab}) stems from the
Lorentz boost from the $\nu N$ CM frame to the laboratory one.
Since $p_T$, $y_H$, and $\phi$ depend on $\phi^\star$, the integration over
$\phi^\star$ in Eq.~(\ref{eq:dif}) is no longer trivial, and we need to insert
the symbolic factor $(1/2\pi)\int_0^{2\pi}d\phi^\star$ on the right-hand side
of that equation.

For future applications, we also present compact formulas that allow us to
determine $p_T^\star$, $y_H^\star$, and $\phi^\star$, once $p_T$, $y_H$, and
$\phi$ are given.
In fact, Eqs.~(\ref{eq:ptlab}) and (\ref{eq:philab}) can be straightforwardly
inverted by observing that the quantity $A$ defined in Eq.~(\ref{eq:a}) can be
expressed in terms of $m_T$ and $y_H$ by substituting Eq.~(\ref{eq:ylab}),
the result being
\begin{equation}
A=\frac{mm_T\exp(y_H)}{W^2+Q^2}\sqrt{\frac{Q^2}{S}(S-W^2-Q^2)}.
\end{equation}
Having obtained $p_T^\star$, we can then evaluate $y_H^\star$ from 
Eq.~(\ref{eq:ylab}).
For the reader's convenience, we collect the relevant formulas here:
\begin{eqnarray}
p_T^\star&=&\sqrt{p_T^2+A(A+2p_T\cos\phi)},
\nonumber\\
y_H^\star&=&y_H+\ln\frac{\sqrt SWm_T}{(W^2+Q^2)m_T^\star}
+\frac{1}{2}\ln\frac{m^2}{S},
\nonumber\\
\cos\phi^\star&=&\frac{p_T\cos\phi+A}{p_T^\star}.
\end{eqnarray}

If the four-momentum $p_H+p^\prime+P-p$ of the hadronic final state $H+j+X$
can be measured in the laboratory frame, then the $Z^\star$ four-momentum can
be evaluated as $q=(p_H+p^\prime+P-p)-P$.
Since the direction of the incident neutrino beam is given by the experimental
set-up, the four-momenta $k$ and $k^\prime$ of the incident and scattered
neutrinos can thus be reconstructed using Eqs.~(\ref{eq:rest}) and
(\ref{eq:res0}).
Then, the $Z^\star N$ CM frame is well-defined, and $p_T^\star$, $y_H^\star$,
and $\phi^\star$ can be determined.
However, if the hadronic final state cannot be fully detected, then one can 
still measure $p_T$, $y_H$, and $\phi$ in the laboratory frame.

We now turn to the cross section of $\nu+N\to\nu+H+X$ in nondiffractive DIS,
which proceeds through the $2\to2$ partonic subprocesses
$\nu+g\to\nu+c\overline{c}[n]$ mentioned in Section~\ref{sec:one}.
We have
\begin{eqnarray}
\frac{d\sigma}{dQ^2}(\nu+N\to\nu+H+X)
&=&\frac{g^4\alpha_s}{6M^3\left(Q^2+M_Z^2\right)^2}\int_{(Q^2+M^2)/S}^1dx\,
f_{g/N}(x,\mu_f)
\nonumber\\
&&{}\times\sum_n\left\langle{\cal O}^H[n]\right\rangle h_n(y,Q^2),
\label{eq:tor}
\end{eqnarray}
where, to LO in $v$, $n={}^1\!S_0^{(8)},{}^3\!S_1^{(8)},{}^3\!P_J^{(8)}$ for
$H=J/\psi,\psi^\prime$ and $n={}^3\!S_1^{(8)}$ for $H=\chi_{cJ}$.
The functions $h_n(y,Q^2)$ may be found in Eq.~(9) of Ref.~\cite{tor}, which
contains a typographical error, as pointed out in Section~\ref{sec:one}.
The kinematically allowed ranges of $S$ and $Q^2$ are $S>M^2$ and
$0<Q^2<S-M^2$, respectively.
The kinematics of $\nu+N\to\nu+H+X$ emerges from that of $\nu+N\to\nu+H+j+X$ 
by nullifying $p^\prime$.
This leads to the simplified relations
$xy=(Q^2+M^2)/S$, $z=1$, $M_{X^\prime}=0$, $\hat s=M^2$, $\hat t=0$,
$\hat u=-Q^2$, $p_T^\star=0$, $y_H^\star=\ln(W/M)$, $p_T=\sqrt{(1-y)Q^2}$,
$y_H=\ln(yS/mm_T)$, and $\cos\phi=-1$.
As before, we have $W^2=yS-Q^2$.
In turn, $Q^2$ and $x$ may be expressed in terms of $p_T$ and $y_H$, as
\begin{equation}
Q^2=\frac{p_T^2}{1-mm_T\exp(y_H)/S},\qquad
x=\frac{m_T\exp(-y_H)/m-M^2/S}{1-mm_T\exp(y_H)/S},
\end{equation}
which allows us to conveniently evaluate the $p_T$ and $y_H$ distributions
from Eq.~(\ref{eq:tor}).

In the remainder of this section, we recall the analysis \cite{kue,pol} of the
total cross section of $\nu+N\to\nu+H+N^\prime$, with
$H=J/\psi,\chi_{c1},\psi^\prime$, in diffractive DIS.
We first consider the case $H=J/\psi$.
The differential cross section of $\nu+N\to\nu+H+N^\prime$ may be be obtained
from the one of $\mu+N\to\mu+H+N^\prime$ in diffractive DIS via a virtual
photon $\gamma^\star$ by adjusting the electroweak couplings and the
propagator, as
\begin{equation}
\frac{d^2\sigma}{dy\,dQ^2}(\nu+N\to\nu+H+N^\prime)=
\left[\frac{g^2v_cQ^2}{e^2e_c\left(Q^2+M_Z^2\right)}\right]^2
\frac{d^2\sigma}{dy\,dQ^2}(\mu+N\to\mu+H+N^\prime).
\end{equation}
The latter can be decomposed as \cite{han}
\begin{eqnarray}
\frac{d^2\sigma}{dy\,dQ^2}(\mu+N\to\mu+H+N^\prime)&=&
\frac{\alpha}{2\pi}\,\frac{y-Q^2/(2Em)}{Q^2(y^2+Q^2/E^2)}
\left\{\left[1+(1-y)^2+\frac{Q^2}{2E^2}\right]\sigma_T(\nu,Q^2)\right.
\nonumber\\
&&{}+\left.\left[2(1-y)-\frac{Q^2}{2E^2}\right]\sigma_L(\nu,Q^2)\right\},
\end{eqnarray}
where $\sigma_T(\nu,Q^2)$ and $\sigma_L(\nu,Q^2)$ are the transverse and 
longitudinal parts of the diffractive cross section of
$\gamma^\star+N\to H+N^\prime$ in virtual photoproduction, respectively.
Here, $E=(S-m^2)/(2m)$ and $\nu=q\cdot P/m=yE=q^0$ are the incoming-lepton and
virtual-photon energies in the nucleon rest frame, respectively, $\sqrt S$ is
the lepton-nucleon CM energy, and the muon mass is neglected.
At small values of $Q^2$, the contribution from longitudinal photons is 
suppressed, and we may put $\sigma_L(\nu,Q^2)=0$.
On the other hand, according to the VDM, the $Q^2$ dependence of
$\sigma_T(\nu,Q^2)$ is approximately described by the familiar
dipole suppression factor, viz.\
\begin{equation}
\sigma_T(\nu,Q^2)=\frac{\sigma(\nu)}{(1+Q^2/M_V^2)^2},
\end{equation}
where $\sigma(\nu)$ is the diffractive cross section of
$\gamma+N\to H+N^\prime$ in real photoproduction and $M_V$ is a
phenomenological mass parameter.
Fits to experimental data of $\gamma^\star+N\to H+N^\prime$ suggest that
$M_V=(3.2\pm0.6)$~GeV \cite{arn}. 
The measured $\nu$ dependence of $\sigma(\nu)$ is well described by the
phenomenological parameterization \cite{kue,pol,cla,wei,dej}
\begin{equation}
\sigma(\nu)=A\exp\frac{-B}{\nu-C},
\label{eq:cla}
\end{equation}
with $A=20$~nb, $B=45$~GeV, and $C=6$~GeV \cite{dej}, which is valid for
$\nu>C$.

We now generalize our considerations to also include the cases
$H=\chi_{c1},\psi^\prime$.
To this end, we recall that the strengths of the $Z$-boson vector and 
axial-vector couplings to the $S$- and $P$-wave charmonia are proportional to
$v_cf_V$ and $a_cf_A$, where \cite{pol}
\begin{eqnarray}
f_V^2&=&\frac{3}{8\pi}\,\frac{|R_S(0)|^2}{m_c^3},
\nonumber\\
f_A^2&=&\frac{9}{4\pi}\,\frac{\left|R_P^\prime(0)\right|^2}{m_c^5},
\end{eqnarray}
respectively.
Here, $R_S(r)$ and $R_P(r)$ denote the radial wave functions of the $S$- and
$P$-wave charmonia, respectively.
Furthermore, we have \cite{bbl}
\begin{eqnarray}
|R_S(0)|^2&=&\frac{2\pi}{9}\,
\left\langle{\cal O}^\psi\left[{}^3\!S_1^{(1)}\right]\right\rangle
\nonumber\\
\left|R_P^\prime(0)\right|^2&=&\frac{2\pi}{9}\,\frac{
\left\langle{\cal O}^{\chi_{cJ}}\left[{}^3\!P_J^{(1)}\right]\right\rangle}
{2J+1}.
\end{eqnarray}
Assuming that
$\sigma(\psi^\prime+N\to\psi^\prime+N^\prime)\approx
\sigma(\chi_{c1}+N\to\chi_{c1}+N^\prime)\approx
\sigma(J/\psi+N\to J/\psi+N^\prime)$, we thus conclude that
$d^2\sigma(\nu+N\to\nu+\psi^\prime+N^\prime)/(dy\,dQ^2)$ and
$d^2\sigma(\nu+N\to\nu+\chi_{c1}+N^\prime)/(dy\,dQ^2)$ may be obtained from
$d^2\sigma(\nu+N\to\nu+J/\psi+N^\prime)/(dy\,dQ^2)$ through multiplication
with the constant factors
\begin{eqnarray}
R_{\psi^\prime}&=&\frac{
\left\langle{\cal O}^{\psi^\prime}\left[{}^3\!S_1^{(1)}\right]\right\rangle}
{\left\langle{\cal O}^{J/\psi}\left[{}^3\!S_1^{(1)}\right]\right\rangle},
\nonumber\\
R_{\chi_{c1}}&=&\frac{a_c^2}{v_c^2}\,\frac{6}{m_c^2}\,\frac{
\left\langle{\cal O}^{\chi_{c0}}\left[{}^3\!P_0^{(1)}\right]\right\rangle}
{\left\langle{\cal O}^{J/\psi}\left[{}^3\!S_1^{(1)}\right]\right\rangle},
\end{eqnarray}
respectively.
Thus, the effective direct-to-prompt conversion factor reads
\begin{equation}
R_{(J/\psi)_{\rm prompt}}=1+R_{\chi_{c1}}B(\chi_{c1}\to J/\psi+\gamma)
+R_{\psi^\prime}B(\psi^\prime\to J/\psi+X).
\end{equation}

The kinematically allowed ranges of $E$, $Q^2$, and $y$ are
\begin{eqnarray}
E&>&M\left(1+\frac{M}{2m}\right),
\nonumber\\
0&<&Q^2<\frac{2E[2Em-M(M+2m)]}{2E+m},
\nonumber\\
\frac{Q^2+M(M+2m)}{2Em}&<&y<1-\frac{Q^2}{4E^2},
\end{eqnarray}
respectively.
Here, the $N^\prime$ mass is taken to be $m$.
In addition, Eq.~(\ref{eq:cla}) implies the cut $y>C/E$.

\section{Numerical results}
\label{sec:three}

We are now in a position to present our numerical results.
We first describe our theoretical input and the kinematic conditions.
We use $m=0.938$~GeV, $m_c=(1.5\pm0.1)$~GeV \cite{pdg}, $M_V=(3.2\pm0.6)$~GeV
\cite{arn}, $M_W=80.419$~GeV, $M_Z=91.1882$~GeV,
$\sin^2\theta_w=1-M_W^2/M_Z^2=0.22225$,
$G_F=1.16639\times10^{-5}$~GeV${}^{-2}$, $\alpha=1/137.036$, and the LO
formula for $\alpha_s^{(n_f)}(\mu)$ with $n_f=3$ active quark flavours
\cite{pdg}.
The CHORUS Collaboration \cite{chorus} chose the target material to be lead
(Pb), which in average consists of $A=207.2$ nucleons, $Z=82$ of them being
protons.
The NOMAD Collaboration \cite{nomad} uses iron (Fe), with $A=55.854$ and 
$Z=26$.
Leaving aside nuclear corrections and appealing to strong-isospin symmetry,
the effective nucleon PDFs may be approximated by
\begin{eqnarray}
f_{a/N}(x,\mu_f)&=&
\frac{1}{A}\left[Zf_{a/p}(x,\mu_f)+(A-Z)f_{b/p}(x,\mu_f)\right],
\nonumber\\
f_{c/N}(x,\mu_f)&=&f_{c/p}(x,\mu_f),
\label{eq:pdf}
\end{eqnarray}
where 
$(a,b)=(u,d),(\overline{u},\overline{d}),(d,u),(\overline{d},\overline{u})$
and $c=s,\overline{s},g$.
We present the cross sections per nucleon, rather than per nucleus.
The dependence of the cross section per nucleus, $\sigma_A$, on the number of
nucleons, $A$, was studied experimentally for $J/\psi$ inclusive
photoproduction \cite{slac,fnal} and muoproduction \cite{emc,nmc}.
The New Muon Collaboration (NMC) \cite{nmc}, who performed the most recent of
these experiments, was able to reconcile the seemingly inconsistent results
of the previous experiments \cite{slac,fnal,emc} by imposing appropriate cuts
on $z$ and $p_T$.
Separating signal events of elastic (coherent) and deep-inelastic (incoherent)
scattering, the $A$ dependences of the respective cross sections $\sigma_A$
were found to differ significantly \cite{fnal,nmc}.
Assuming that $\sigma_A$ is related to the cross section per free nucleon 
(hydrogen), $\sigma_1$, by the simple power law $\sigma_A=A^\alpha\sigma_1$
\cite{fnal,e706}, with $\alpha$ being an $A$-independent number, we extract
the values $\alpha_{\rm coh}=1.19\pm0.02$ and $\alpha_{\rm in}=1.05\pm0.03$
from the NMC results for coherent and incoherent scattering, respectively.
For lead (iron) targets, this implies that the elastic and deep-inelastic
cross sections per (bound) nucleon, $\sigma_A/A$, are enhanced relative to the
respective values of $\sigma_1$ by factors of $2.75{+0.32\atop-0.27}$
($2.15{+0.18\atop-0.17}$) and $1.31{+0.22\atop-0.20}$
($1.22{+0.16\atop-0.14}$), respectively.
We note in passing that the above value of $\alpha_{\rm in}$ is in same ball
park as those recently found for inclusive large-$p_T$ prompt-photon and
neutral-pion hadroproduction in fixed-target scattering, namely 1.04 and 1.08,
respectively \cite{e706}.
We assume that the power laws of $J/\psi$ muoproduction, which dominantly
proceeds via the electromagnetic current, carry over to $\nu N$ NC DIS and 
employ them to determine the nuclear corrections.
As for the proton PDFs, we employ the LO set by Martin, Roberts, Stirling, and 
Thorne (MRST98LO) \cite{mrst}, with asymptotic scale parameter
$\Lambda^{(4)}=174$~MeV, as our default and the LO set by the CTEQ
Collaboration (CTEQ5L) \cite{cteq}, with $\Lambda^{(4)}=192$~MeV, for
comparison.
The corresponding values of $\Lambda^{(3)}$ are 204~MeV and 224~MeV, 
respectively.
We choose the renormalization and factorization scales to be
$\mu_i=\xi_i\sqrt{Q^2+M^2}$, with $i=r,f$, respectively, and independently
vary the scale parameters $\xi_r$ and $\xi_f$ between 1/2 and 2 about the
default value 1.
We adopt the NRQCD MEs from Table~I of Ref.~\cite{bkl}, which were determined
there from the measured leptonic annihilation rates of the $J/\psi$ and
$\psi^\prime$ mesons and fits to the Tevatron data \cite{abe}.
As for the $J/\psi$ and $\psi^\prime$ mesons, the fit results for
$\left\langle{\cal O}^\psi\left[{}^1\!S_0^{(8)}\right]\right\rangle$ and
$\left\langle{\cal O}^\psi\left[{}^3\!P_0^{(8)}\right]\right\rangle$ are
strongly correlated, so that the linear combinations
\begin{equation}
M_r^\psi=\left\langle{\cal O}^\psi\left[{}^1\!S_0^{(8)}\right]\right\rangle
+\frac{r}{m_c^2}
\left\langle{\cal O}^\psi\left[{}^3\!P_0^{(8)}\right]\right\rangle,
\label{eq:mr}
\end{equation}
with suitable values of $r$, are quoted.
Since Eq.~(\ref{eq:dif}) is sensitive to different linear combinations of
$\left\langle{\cal O}^\psi\left[{}^1\!S_0^{(8)}\right]\right\rangle$ and
$\left\langle{\cal O}^\psi\left[{}^3\!P_0^{(8)}\right]\right\rangle$ than 
appear in Eq.~(\ref{eq:mr}), we write
$\left\langle{\cal O}^\psi\left[{}^1\!S_0^{(8)}\right]\right\rangle
=\kappa_\psi M_r^\psi$
and
$\left\langle{\cal O}^\psi\left[{}^3\!P_0^{(8)}\right]\right\rangle
=(1-\kappa_\psi)\left(m_c^2/r\right)M_r^\psi$ and independently vary
$\kappa_{J/\psi}$ and $\kappa_{\psi^\prime}$ between 0 and 1 around the
default value 1/2.
The wide-band neutrino beam utilized in the CHORUS and NOMAD experiments
mainly consists of $\nu_\mu$ neutrinos and contains $\overline{\nu}_\mu$,
$\nu_e$, and $\overline{\nu}_e$ admixtures of about 6\%, 1\%, and below 1\%,
respectively.
The average $\nu_\mu$ energy is 27~GeV.
Using accurate parameterizations for the energy spectra $\phi_\nu(E_\nu)$ of
the individual neutrino species $\nu$ \cite{nomad}, we calculate the 
spectrum-averaged cross section as
\begin{equation}
\sigma=\frac{\sum_\nu\int dE_\nu\,\phi_\nu(E_\nu)\sigma_\nu(E_\nu)}
{\sum_\nu\int dE_\nu\,\phi_\nu(E_\nu)},
\end{equation}
where $\nu=\nu_e,\overline{\nu}_e,\nu_\mu,\overline{\nu}_\mu$.
In order to conform with the CHORUS analysis, we apply the acceptance cut
$20<E_\nu<200$~GeV.

In order to estimate the theoretical uncertainties in our predictions, we vary
the unphysical parameters $\xi_r$, $\xi_f$, $\kappa_{J/\psi}$, and
$\kappa_{\psi^\prime}$ as indicated above, take into account the experimental
errors on $\alpha_{\rm coh}$, $\alpha_{\rm in}$, $m_c$, $M_V$, the relevant
charmonium decay branching fractions, and the default MEs, and switch from our
default PDF set to the CTEQ5L one, properly adjusting $\Lambda^{(3)}$ and the
MEs.
We then combine the individual shifts in quadrature, allowing for the upper 
and lower half-errors to be different.

In Figs.~\ref{fig:pT} and \ref{fig:y}, we respectively present the $p_T$ and
$y_{J/\psi}$ distributions of prompt $J/\psi$ production in nondiffractive
$\nu N$ NC DIS under CHORUS experimental conditions predicted to LO in NRQCD.
Specifically, the $2\to2$ CO contributions (upper histograms) are compared 
with the $2\to3$ CS ones (lower histograms).
In each case, the theoretical errors, evaluated as explained above, are 
indicated by the hatched areas.
Since the contributing partonic cross sections are all gluon initiated, our
our central predictions can be rendered appropriate for NOMAD by simply
adjusting the nuclear correction factor, by multiplication with 0.94.
We observe that the shapes of the CO and CS distributions are very similar.
However, the CO distributions greatly exceed the CS ones in normalization.
In fact, the integrated CO and CS cross sections are
$3.0{+1.6\atop-1.7}\times10^{-3}$~fb and $1.0{+0.7\atop-0.5}\times10^{-4}$~fb,
respectively, the ratio of the central values being 30.
The bulk parts, 72\% in the CO case and 78\% in the CS case, are due to direct
$J/\psi$ production.
In the first case, the fractions due to feed-down from $\chi_{cJ}$ and
$\psi^\prime$ mesons are each 14\%, while, in the CS case, there is only feed
down from $\psi^\prime$ mesons.
Therefore, at first sight, the suppression of the CSM prediction seems to
contradict the na\"\i ve expectation based on the NRQCD counting rules
(see Table~\ref{tab:one}).
However, detailed inspection reveals that this suppression can be attributed
to the smallness of the effective value of $\sqrt S$, which affects the
$2\to3$ phase space of the CS process more severely than the $2\to2$ one of
the CO processes.
For $E=27$~GeV, we just have $\sqrt S=7.2$~GeV.
For a monoenergetic neutrino beam with $E=48$~TeV, so that $\sqrt S=300$~GeV
as for $ep$ collisions at DESY HERA, the CO to CS ratio becomes 4.5.
This should be compared with the corresponding analysis for direct $J/\psi$ 
production in nondiffractive $ep$ DIS at HERA, with $Q^2>4$~GeV${}^2$, which
yields a CO to CS ratio of 7.8 \cite{ep}.

We find the total cross section of prompt $J/\psi$ production in diffractive
$\nu N$ NC DIS to be $3.3{+1.6\atop-1.4}\times10^{-3}$~fb.
The theoretical uncertainty quoted here is merely parametric; it neither
includes the experimental error on $\sigma(\nu)$ as parameterized by
Eq.~(\ref{eq:cla}), nor does it reflect the unavoidable lack of rigour of the
underlying model assumptions.
The fractions due to feed-down from $\chi_{c1}$ and $\psi^\prime$ mesons 
amount to 19\% and 17\%, respectively.
The corresponding prediction for NOMAD is obtained through multiplication with
the scaling factor 0.80, which accounts for the difference in the nuclear
correction factor.

The CHORUS Collaboration measured the spectrum-averaged total cross section 
per nucleon of $\nu+N\to\nu+J/\psi+X$ in DIS to be
$(6.3\pm3.0)\times10^{-2}$~fb.
In Fig.~\ref{fig:exp}, this experimental data point is shown together with our
theoretical predictions for the CO, CS, and diffractive contributions.
It should be compared with their sum, $6.4\times10^{-3}$~fb.
As for the central values, the measurement exceeds the combined prediction by
a factor of 10.
We have to bear in mind that our theoretical prediction for the nondiffractive
cross section is of LO in $\alpha_s$ and $v$.
The NLO corrections are expected to lead to a substantial enhancement.
It would be desirable if the CHORUS Collaboration were able to substantially
increase their data sample and to exclude the diffractive regime by imposing
suitable acceptance cuts on $z$, $p_T^\star$, or $M_{X^\prime}$, and if the
NOMAD and NuTeV Collaborations could come up with similar measurements.

\section{Conclusions}
\label{sec:four}

We provided, in analytic form, the cross sections of the partonic subprocesses
$\nu+a\to\nu+c\overline{c}[n]+a$, where $a=q,\overline{q},g$ and
$n={}^3\!S_1^{(1)},{}^3\!P_J^{(1)},{}^1\!S_0^{(8)},{}^3\!S_1^{(8)},
{}^3\!P_J^{(8)}$, to LO in the NRQCD factorization formalism \cite{bbl}.
We also confirmed previous results for the cross sections of the partonic 
subprocesses $\nu+g\to\nu+c\overline{c}[n]$, where
$n={}^1\!S_0^{(8)},{}^3\!S_1^{(8)},{}^3\!P_J^{(8)}$ \cite{tor}.
Using these results, we then studied the cross section of prompt $J/\psi$ 
inclusive production in nondiffractive $\nu N$ NC DIS to LO in NRQCD and the
CSM.
We presented the cross section per nucleon bound in lead and averaged it over
the effective energy spectrum of the wide-band neutrino beam employed in the
CHORUS \cite{chorus} and NOMAD \cite{nomad} experiments.
Apart from the total cross section, we also considered the $p_T$ and 
$y_{J/\psi}$ distributions.
The cross sections of the $2\to3$ CO partonic subprocesses, which did not 
enter our phenomenological study, can be used in the future, at LO, in
connection with an appropriate acceptance cut on $z$, $p_T^\star$, or
$M_{X^\prime}$ to exclude the collinear singularities in the limit
$\hat t\to0$ or, at NLO, as an essential ingredient for the real radiative
corrections to the inclusive cross section of $\nu+N\to\nu+J/\psi+X$.
A similar comment applies to the $2\to3$ CS partonic subprocesses of
$\chi_{cJ}$ production, where the singularities are of IR type.
Repeating the analysis of Refs.~\cite{kue,pol} with up-to-date input, we also
evaluated the total cross section of prompt $J/\psi$ inclusive production in
diffractive $\nu N$ NC DIS.

As for nondiffractive direct $J/\psi$ production, from the NRQCD counting
rules, one expects the CO contribution, which arises from $2\to2$ partonic
subprocesses, to be slightly suppressed, by a factor of
$v^4/\alpha_s\approx\alpha_s$ relative to the CS one, which is generated by
$2\to3$ partonic subprocesses.
However, our analysis revealed that, under CHORUS kinematic conditions, the CS
contribution is approximately 30 times smaller than the CO one.
We demonstrated that this may be attributed to the smallness of the effective
value of $\sqrt S$, which constrains the $2\to3$ phase space more severely
than the $2\to2$ one.

As far as central values are concerned, the CHORUS measurement of the
spectrum-averaged total cross section per nucleon of $\nu+N\to\nu+J/\psi+X$ in
DIS overshoots our combined prediction by a factor of 10.
However, this comparison does not yet permit any meaningful conclusions 
concerning the validity of the NRQCD factorization formalism \cite{bbl}.
On the one hand, our theoretical prediction for the nondiffractive 
contribution should be substantially increased by the inclusion of NLO
corrections, which are presently still unknown.
On the other hand, the experimental error is still rather sizeable, and the
diffractive events have not been separated from the experimental data set.

\bigskip
\noindent
{\bf Acknowledgements}
\smallskip

\noindent
We are indebted to Maarten de Jong and Lalit Sehgal for useful communications
concerning Refs.~\cite{dej} and \cite{seh}, respectively, and to Kai Zuber for
providing us with the energy spectra of the CERN wide-band neutrino beam in
numerical form and for helpful discussions regarding Ref.~\cite{nomad}.
The work of L.Z. was supported by the Studienstiftung des deutschen Volkes
through a PhD scholarship.
This work was supported in part by the Deutsche Forschungsgemeinschaft through
Grant No.\ KN~365/1-1, by the Bundesministerium f\"ur Bildung und Forschung
through Grant No.\ 05~HT1GUA~4, by the European Commission through the
Research Training Network {\it Quantum Chromodynamics and the Deep Structure
of Elementary Particles} under Contract No.\ ERBFMRX-CT98-0194, and by Sun
Microsystems through Academic Equipment Grant No.~EDUD-7832-000332-GER.

\def\theequation{\Alph{section}.\arabic{equation}}
\begin{appendix}
\setcounter{equation}{0}
\section{Partonic cross sections}

In this Appendix, we present analytic expressions for the coefficients
$F_a[n]$, $T_a[n]$, $L_a[n]$, and $A_a[n]$ appearing in Eq.~(\ref{eq:res}).
In order to compactify the expressions, it is useful to introduce the Lorentz
invariants $s=2q\cdot p$, $t=-2p\cdot p^\prime$, and $u=-2q\cdot p^\prime$,
which are related to the partonic Mandelstam variables by $s=\hat s+Q^2$,
$t=\hat t$, and $u=\hat u+Q^2$, respectively.

$\nu(\overline{\nu})+q(\overline{q})\to\nu(\overline{\nu})
+c\overline{c}\left[{}^3\!S_1^{(1)}\right]+q(\overline{q})$:
\begin{equation}
F=T=L=A=0.
\end{equation}

$\nu(\overline{\nu})+q(\overline{q})\to\nu(\overline{\nu})
+c\overline{c}\left[{}^3\!P_J^{(1)}\right]+q(\overline{q})$:
\begin{equation}
F=T=L=A=0.
\end{equation}

$\nu(\overline{\nu})+q(\overline{q})\to\nu(\overline{\nu})
+c\overline{c}\left[{}^1\!S_0^{(8)}\right]+q(\overline{q})$:
\begin{eqnarray}
F&=&\frac{-g^4\alpha_s^2v_c^2Q^2}
{6 \pi M (Q^2 + M_Z^2)^2 s^4 t (s + u)^2},
\nonumber\\
T&=&2 Q^4 t^2 + 2 Q^2 s t ( s + u ) + s^4 + s^2 u^2,
\nonumber\\
L&=&-2 Q^2 t (Q^2 t + s u),
\nonumber\\
A&=&0.
\end{eqnarray}

$\nu(\overline{\nu})+q(\overline{q})\to\nu(\overline{\nu})
+c\overline{c}\left[{}^3\!S_1^{(8)}\right]+q(\overline{q})$:
\begin{eqnarray}
F&=&\frac{g^4 \alpha_s^2}{36\pi M^3 
(Q^2 + M_Z^2)^2 (Q^2 - s)^2 (Q^2 - u)^2 s^4 t (s + u)^2},
\nonumber\\
T&=& -(a_q^2 + v_q^2) Q^2 t (s + u)^2 \{2 Q^6 t^2 (2 s + t) 
     + 2 Q^4 s [s^2 u - s t (3 t - 2 u) -2 t^3]
     + Q^2 s^2 [s^2 (t
\nonumber\\
 & & {}- 2 u) + 2 s (t^2 - 4 t u - u^2) + 
     t (2 t^2 - 2 t u - u^2)]
     + s^3 u (s^2 + 2 s t + 2 t^2 + 2 t u + u^2)\}
\nonumber\\
 & & {}+ 4 a_q a_c Q^2 t (s + u) (Q^2 - s) (Q^2 - u) 
     \{Q^6 t (s + 2 t) + Q^4 [s^3 - s^2 (t - 2 u) - 5 s t^2 
     - 2 t^2 (t
\nonumber\\
 & & {}+ u)] - Q^2 s [s^3 + s^2 (t + 5 u) 
     - s (2 t^2 - 5 t u - 3 u^2)
     - t (2 t^2 + t u - u^2)]  
     + s^2 u [2 s^2 + s (4 t
\nonumber\\
 & & {}+ 3 u) + 2 t^2  + 3 t u + u^2]\}
     - 2 a_c^2 Q^2 (Q^2 - s)^2 (Q^2 - u)^2 
     \{2 Q^6 t^2 + 2 Q^4 t [s^2 - s (3 t - u) - 3 t (t
\nonumber\\
 & & {}+ u)] - Q^2 [s^4 + 6 s^3 t + s^2 (2 t^2 + 12 t u 
     + u^2) - 2 s t (4 t^2 + t u - 3 u^2) - 4 t^2 (t + u)^2]
     + s^5
\nonumber\\
 & & {}+ s^4 (5 t + u) + s^3 (8 t^2 + 12 t u + u^2)
     + s^2 (4 t^3 + 16 t^2 u + 13 t u^2 + u^3)
     + 4 s t u (t + u)^2\} ,
\nonumber\\
L&=& 2 (a_q^2 + v_q^2) Q^4 (Q^2 - s)^2 t^2 (s + t)^2 (s + u)^2 
     - 4 a_q a_c Q^2 (Q^2 - s)^2 (Q^2 - u) t (s + u) [Q^4 t (s
\nonumber\\
 & & {}+ 2 t) - 
     Q^2 (s^3 + 2 s^2 t + 3 s t^2 + 2 t^3 + 2 t^2 u) + 
     s^4 + s^3 (2 t + u) + s^2 t^2 - s t u (t + u) ] 
     + 2 a_c^2 (Q^2
\nonumber\\
 & & {}- s)^2 (Q^2 - u)^2 \{2 Q^8 t^2 
     - 2 Q^6 t [s^2 + s (3 t - u) + 3 t (t + u)]
     + Q^4 [s^4 + 4 s^3 t + s^2 (8 t^2 - 2 t u
\nonumber\\
 & & {}+ u^2) + 2 s t (4 t^2 + t u 
     - 3 u^2)
     + 4 t^2 (t + u)^2]
     - 2 Q^2 s [s^4 + s^3 (2 t + u) + s^2 (2 t^2 + u^2)
     + s (t^3
\nonumber\\
 & & {}- 2 t^2 u - 2 t u^2 + u^3)
     - 2 t u (t + u)^2]
     + s^2 (s + t + u)^2 (s^2 + u^2)\},
\nonumber\\
A&=& \pm 2 v_q Q^2 s t (s + u) \{
     a_q (s + u) [2 Q^6 s t - 2 Q^4 t (2 s^2 + s t - t u)
\nonumber\\
&&{}+ Q^2 s (s^2 (t - 2 u) + 2 s (t^2 + u^2) - t u (2 t - u))
     + s^2 u (s - u) (s + 2 t + u)]
\nonumber\\
&&{}-2 a_c (Q^2 - s) (Q^2 - u) 
     [Q^6 t - Q^4 (s^2 + 3 s t + t^2) 
\nonumber\\
&&{}+ Q^2 (s^3 + s^2 (3 t + u) + s(2 t^2 + t u + u^2) - t u (t + u))
     - s u^2 (s + t + u)] \},
\end{eqnarray}
where the plus (minus) sign refers to a $\nu+q$ or 
$\overline{\nu}+\overline{q}$ ($\nu+\overline{q}$ or $\overline{\nu}+q$)
initial state.

$\nu(\overline{\nu})+q(\overline{q})\to\nu(\overline{\nu})
+c\overline{c}\left[{}^3\!P_J^{(8)}\right]+q(\overline{q})$:
\begin{eqnarray}
F&=&\frac{2g^4\alpha_s^2v_c^2Q^2}{3 \pi M^3
(Q^2 + M_Z^2)^2 s^4 t (s + u)^4},
\nonumber\\
T&=& 8 Q^6 t [2 s^2 + s t + t (2 t + u)] 
     - 2 Q^4 [4 s^4 + 12 s^3 t + s^2 (19 t^2 + 8 t u + 4 u^2)
     + 2 s t (6 t^2 
\nonumber\\
 & & {}+ t u - 2 u^2) + t^2 (8 t^2 + 12 t u + 7 u^2)] 
     + 2 Q^2 s [6 s^4 + s^3 (7 t + 6 u) + s^2 (4 t^2 + 3 t u
\nonumber\\
 & & {}+ 6 u^2) 
     - s u ( 8 t^2 + 3 t u - 6 u^2) - t u (8 t^2 + 12 t u + 7 u^2)] 
     - s^2 (s + u) [7 s^3 + s^2 (12 t
\nonumber\\
 & & {}+ 7 u) + s (8 t^2 + 16 t u + 7 u^2)
     + u (8 t^2 + 12 t u + 7 u^2)],
\nonumber\\
L&=& 8 Q^6 t [s^2 - s t - t (2 t + u)] 
     -2 Q^4 [2 s^4 + 4 s^3 t + s^2 (5 t^2 + 8 t u + 2 u^2)
     - 2 s t (6 t^2
\nonumber\\
 & & {}+ t u - 2 u^2) - t^2 (8 t^2 + 12 t u + 7 u^2)]
     + 2 Q^2 s (s + u) [2 s^3 + 2 s^2 t + s (8 t^2 + 5 t u
\nonumber\\
 & & {} + 2 u^2)+ t (8 t^2 + 12 t u + 7 u^2)],
\nonumber\\
A&=&0.
\end{eqnarray}

$\nu(\overline{\nu})+g\to\nu(\overline{\nu})
+c\overline{c}\left[{}^3\!S_1^{(1)}\right]+g$:
\begin{eqnarray}
F&=&\frac{-2g^4\alpha_s^2v_c^2Q^2}{27\pi 
M (Q^2 + M_Z^2)^2 s^4 (s + t)^2 (s + u)^2 (t + u)^2},
\nonumber\\
T&=& 4 Q^6 t^2 (s^2 + t^2) 
     - 2 Q^4 t [s^3 (3 t - 2 u) + 3 s^2 t (t + u) + 2 s t^2 (t - u)
     + 2 t^3 (t + u)]
\nonumber\\ 
 & & {}+ 2 Q^2 s [s^3 (t - u)^2 - 2 s^2 t u (t + u)
     - s t^2 u (2 t - u) - 2 t^3 u (t + u)] 
     - 2 s^2 [s^3 (t^2
\nonumber\\
 & & {}+ t u + u^2) + s^2 (t + u)^3 + s t u (t^2 + 3 t u + u^2)
     +t^2 u^2 (t + u)],
\nonumber\\
L&=& 2 Q^6 t^2 (s^2 - 2 t^2) 
     - 2 Q^4 t (s^2 - 2 t^2) 
     [s (t - u) + t (t + u)] 
     + Q^2 s [s^3 (t^2 + u^2)
\nonumber\\
 & & {}+ 2 s^2 t^2 (t + u) + s t^2 (t^2 + 6 t u + u^2)
     + 4 t^3 u (t + u)],
\nonumber\\
A&=&0.
\end{eqnarray}

$\nu(\overline{\nu})+g\to\nu(\overline{\nu})
+c\overline{c}\left[{}^3\!P_0^{(1)}\right]+g$:
\begin{eqnarray}
F&=&\frac{8 g^4 \alpha_s^2 a_c^2 (Q^2+M^2)^2}{27 \pi M^3 (Q^2 + M_Z^2)^2 
s^4 (s + t)^4 (s + u)^4 (t + u)^4},
\nonumber\\
T&=& 4 Q^8 t^2 [s^6 - 2 s^4 u^2+ 
     s^3 (t^3 - t u^2) + s^2 (2 t^4 + 4 t^3 u + 2 t^2 u^2 + u^4) + 
     s t (2 t^4 + 6 t^3 u + 5 t^2 u^2 \nonumber\\
 & & + 2 t u^3 + u^4)+ t^2 (t + u)^2 (t^2 + u^2)]
     - 4 Q^6 t [s^7 (2 t - u) + s^6 t (t + u)- 
     2 s^5 (t^3 + t^2 u + 2 t u^2\nonumber\\
 & & - u^3) - 
     s^4 t (t^3 + 6 t^2 u + 7 t u^2 + 2 u^3) + 
     s^3 (t^5 - 3 t^4 u - 7 t^3 u^2 - 3 t^2 u^3 + 2 t u^4 - u^5) + 
     s^2 t (t^5\nonumber\\
 & & + 4 t^4 u + 3 t^3 u^2 + 2 t^2 u^3 + 3 t u^4 + u^5)+ 
     s t^2 u (t + u)^2 (2 t^2 + 5 t u + u^2)+ 2 t^3 u^2 (t + u)^3]
     \nonumber\\ 
 & & +2 Q^4 t [2 s^8 (t - 2 u)+ 2 s^7 (t^2 - u^2)- 
     s^6 (3 t^3 - 2 t^2 u - t u^2 - 8 u^3)- 
     2 s^5 (2 t^4 + 4 t^3 u + t^2 u^2\nonumber\\
 & & - 4 t u^3 - 3 u^4)+ 
     s^4 (t^5 - 6 t^4 u - 6 t^3 u^2 + 2 t^2 u^3 + 3 t u^4 - 4 u^5)+ 
     2 s^3 (t + u)^2 (2 t^4 + 2 t^3 u\nonumber\\
 & & - t^2 u^2 + t u^3 - 2 u^4)+ 
     s^2 t (t + u)^2 (2 t^4 + 12 t^3 u + 9 t^2 u^2 - 4 t u^3 - 2 u^4)+ 
     2 s t^2 u (t + u)^3 (2 t^2\nonumber\\
 & & + 4 t u - u^2)+ 2 t^3 u^2 (t + u)^4]
     + 2 Q^2 s t u [2 s^8 + 2 s^7 (t + u)- s^6 (3 t^2 + 2 t u + 3 u^2)- 
     2 s^5 (2 t^3\nonumber\\
 & & + 5 t^2 u + 6 t u^2 + 3 u^3)+ 
     s^4 (t^4 - 4 t^3 u - 8 t^2 u^2 - 4 t u^3 + u^4)+ 
     2 s^3 (2 t^5 + 7 t^4 u + 11 t^3 u^2\nonumber\\
 & & + 10 t^2 u^3 + 6 t u^4 + 2 u^5)+ 
     s^2 (t + u)^2 (2 t^4 + 12 t^3 u + 15 t^2 u^2 + 6 t u^3 + 2 u^4)+ 
     2 s t u (t + u)^3 (2 t^2\nonumber\\
 & & + 4 t u + u^2)+ 2 t^2 u^2 (t + u)^4],
\nonumber\\
L&=&-2 Q^8 t^2 [s^6 - 2 s^4 u^2+ 
     2 s^3 (t^3 - t u^2) + s^2 (4 t^4 + 8 t^3 u + 4 t^2 u^2 + u^4) + 
     2 s t (2 t^4 + 6 t^3 u + 5 t^2 u^2\nonumber\\
 & & + 2 t u^3 + u^4)
     + 2 t^2 (t + u)^2 (t^2 + u^2)]+ 
     2 Q^6 t [s^7 (3 t - u) + 3 s^6 t (t + u)- 
     2 s^5 (2 t^3 + 3 t^2 u\nonumber\\
 & & + 4 t u^2 - u^3)- 
     2 s^4 t (t^3 + 8 t^2 u + 9 t u^2 + 2 u^3)+ 
     s^3 (6 t^5 - 2 t^4 u - 18 t^3 u^2 - 8 t^2 u^3 + 5 t u^4 - u^5)
     \nonumber\\
 & & + s^2 t (4 t^5 + 12 t^4 u + 6 t^3 u^2 + 4 t^2 u^3 + 7 t u^4 + u^5)+ 
     2 s t^2 u (t + u)^2 (2 t^2 + 5 t u + u^2)+ 4 t^3 u^2 (t\nonumber\\
 & & + u)^3]
     -Q^4 [s^8 (7 t^2 - 4 t u + u^2)+ 
     2 s^7 t (7 t^2 + 5 t u - 2 u^2)- 
     s^6 (5 t^4 - 6 t^3 u - t^2 u^2 - 12 t u^3\nonumber\\
 & & + 2 u^4)
     -8 s^5 t (3 t - u) (t + u)^3- s^4 (8 t^6 + 76 t^5 u + 110 t^4 u^2 
     + 40 t^3 u^3 - 5 t^2 u^4 + 8 t u^5 - u^6)\nonumber\\
 & & +2 s^3 t (4 t^6 + 2 t^5 u - 22 t^4 u^2 - 22 t^3 u^3 - t^2 u^4 - t u^5 - 
       2 u^6)+ 
     s^2 t^2 (t + u)^2 (4 t^4 + 24 t^3 u + 8 t^2 u^2\nonumber\\
 & & - 4 t u^3 - u^4)
     + 4 s t^3 u (t + u)^3 (2 t^2 + 4 t u - u^2)
     + 4 t^4 u^2 (t + u)^4]
     + 2 Q^2 s (s + u) [s^7 (2 t^2 - t u\nonumber\\
 & & + u^2)+ 2 s^6 t^2 (3 t + u)+ 
     s^5 (4 t^4 + 9 t^3 u + 6 t^2 u^2 + 3 t u^3 - 2 u^4)- 
     s^4 t (4 t^4 - 11 t^2 u^2 - 12 t u^3\nonumber\\
 & & - u^4)- 
     s^3 (6 t^6 + 18 t^5 u + 12 t^4 u^2 - t^3 u^3 + 2 t^2 u^4 + 2 t u^5 - 
       u^6)- 
     s^2 t (t + u)^2 (2 t^4 + 12 t^3 u\nonumber\\
 & & + 6 t^2 u^2 + 2 t u^3 + u^4)
     - 2 s t^3 u (t + u)^3 (2 t + 3 u)-2 t^3 u^2 (t + u)^4]
     -s^2 (s + u)^2 (s + t + u)^2 [s^4\nonumber\\
 & & (t^2 + u^2)+ 2 s^3 (t^3 - u^3)+ 
     s^2 (t^4 + 2 t^3 u + 2 t^2 u^2 + u^4)+2 s t^3 u (t + u)+ 
     t^2 u^2 (t + u)^2],
\nonumber\\
A&=&0.
\end{eqnarray}

$\nu(\overline{\nu})+g\to\nu(\overline{\nu})
+c\overline{c}\left[{}^3\!P_1^{(1)}\right]+g$:
\begin{eqnarray}
F&=&\frac{4 g^4 \alpha_s^2 a_c^2}{27 \pi M^3 (Q^2 + M_Z^2)^2 s^4 (s + t)^4 
    (s + u)^4 (t + u)^4},
\nonumber\\
T&=& -8 Q^{10} t^4 (s + t) (t + u) [3 s^3+ 
     s^2 (7 t + 10 u) + s (5 t^2 + 15 t u + 9 u^2) 
     + t^3 + 5 t^2 u + 7 t u^2 \nonumber\\
 & & + 2 u^3] -4 Q^8 t^2 [s^8 + 7 s^7 (t + u) + s^6 (24 t^2 + 48 t u + 22 u^2) +
     2 s^5 (17 t^3 + 62 t^2 u + 60 t u^2 \nonumber\\
 & & + 15 u^3) + s^4 (7 t^4 + 112 t^3 u + 219 t^2 u^2 + 130 t u^3 + 17 u^4) - 
     s^3 (30 t^5 + 37 t^4 u - 91 t^3 u^2 \nonumber\\
 & & - 151 t^2 u^3 - 56 t u^4 - 3 u^5)- 
     s^2 t (t + u)^2 (31 t^3 + 62 t^2 u - 18 t u^2 - 4 u^3)- 
     s t (11 t^6 + 69 t^5 u\nonumber\\
 & & + 144 t^4 u^2 + 123 t^3 u^3 + 42 t^2 u^4 +
     6 t u^5 + u^6)- 
     t^3 (t + u)^2 (t^3 + 9 t^2 u + 17 t u^2 + 7 u^3)]\nonumber\\
 & & + 2 Q^6 t [s^9 (4 t - 2 u) + 2 s^8 (16 t^2 + 9 t u - 7 u^2) + 
     s^7 (125 t^3 + 206 t^2 u + 25 t u^2 - 44 u^3)\nonumber\\
 & & + s^6 (239 t^4 + 695 t^3 u + 507 t^2 u^2 - 9 t u^3 - 60 u^4) + 
     s^5 (221 t^5 + 1022 t^4 u + 1410 t^3 u^2\nonumber\\
 & & + 602 t^2 u^3 - 35 t u^4 - 
     34 u^5)+ s^4 (91 t^6 + 689 t^5 u + 1548 t^4 u^2 + 
     1318 t^3 u^3 + 361 t^2 u^4\nonumber\\
 & & - 13 t u^5 - 6 u^6)+ 
     s^3 t (t + u)^2 (16 t^4 + 156 t^3 u + 357 t^2 u^2 + 78 t u^3 + 2 u^4)+ 
     s^2 t^2 (10 t^6\nonumber\\
 & & + 24 t^5 u + 57 t^4 u^2 + 127 t^3 u^3 + 99 t^2 u^4 - 
     t u^5 - 16 u^6)+ 2 s t^2 (t + u)^2 (4 t^5 + 2 t^4 u\nonumber\\
 & & - 17 t^3 u^2 - 
     19 t^2 u^3 - 12 t u^4 - 4 u^5)+ 
     2 t^3 (t + u)^4 (t^3 - 4 t u^2 - u^3)] 
     - 2 Q^4 [s^{10} (3 t^2\nonumber\\
 & & - 2 t u + u^2)+ 
     s^9 (25 t^3 + 7 t^2 u - 11 t u^2 + 7 u^3)+ 
     s^8 (100 t^4 + 132 t^3 u - 25 t^2 u^2 - 22 t u^3\nonumber\\
 & & + 23 u^4)+ 
     s^7 (208 t^5 + 539 t^4 u + 277 t^3 u^2 - 93 t^2 u^3 - 5 t u^4 + 34 u^5)
     + s^6 (217 t^6\nonumber\\
 & & + 935 t^5 u + 1162 t^4 u^2 + 
     366 t^3 u^3 - 66 t^2 u^4 + 29 t u^5 + 23 u^6)+ 
     s^5 (87 t^7 + 688 t^6 u\nonumber\\
 & & + 1545 t^5 u^2 + 1320 t^4 u^3 + 406 t^3 u^4 + 
     59 t^2 u^5 + 36 t u^6 + 7 u^7)- s^4 (t + u)^2 
     (22 t^6 - 119 t^5 u\nonumber\\
 & & - 396 t^4 u^2 - 184 t^3 u^3 - 72 t^2 u^4 - 
     17 t u^5 - u^6)- 
     s^3 t (28 t^8 + 152 t^7 u + 257 t^6 u^2 + 100 t^5 u^3\nonumber\\
 & & - 163 t^4 u^4 - 
     241 t^3 u^5 - 158 t^2 u^6 - 51 t u^7 - 4 u^8)- 
     s^2 t^2 (t + u)^2 (6 t^6 + 52 t^5 u + 132 t^4 u^2\nonumber\\
 & & + 125 t^3 u^3 + 
     47 t^2 u^4 - 5 t u^5 - 7 u^6)- 2 s t^3 u (t + u)^4 
     (3 t^3 + 12 t^2 u + 8 t u^2 + 2 u^3)-2 t^4 u^2 (t\nonumber\\
 & & + u)^6]+ 
     2 Q^2 s [s^{10} (t^2 + u^2)+ 
     s^9 (8 t^3 + 6 t^2 u + 6 t u^2 + 8 u^3) + 
     s^8 (32 t^4 + 56 t^3 u + 42 t^2 u^2\nonumber\\
 & & + 52 t u^3 + 30 u^4) + 
     s^7 (75 t^5 + 232 t^4 u + 261 t^3 u^2 + 221 t^2 u^3 + 174 t u^4 + 
     57 u^5)+ 
     s^6 (105 t^6\nonumber\\
 & & + 508 t^5 u + 888 t^4 u^2 + 836 t^3 u^3 + 576 t^2 u^4 + 
     280 t u^5 + 57 u^6)+ 2 s^5 (43 t^7 + 308 t^6 u + 791 t^5 u^2\nonumber\\
 & & + 1015 t^4 u^3 + 774 t^3 u^4 + 389 t^2 u^5 + 119 t u^6 + 15 u^7)+ 
     s^4 (t + u)^2 (38 t^6 + 332 t^5 u + 799 t^4 u^2\nonumber\\
 & & + 754 t^3 u^3 + 
     389 t^2 u^4 + 96 t u^5 + 8 u^6) + 
     s^3 (7 t^9 + 134 t^8 u + 734 t^7 u^2 + 1847 t^6 u^3 + 2546 t^5
     u^4\nonumber\\
 & & + 2066 t^4 u^5 + 993 t^3 u^6 + 262 t^2 u^7 + 30 t u^8 + u^9)
     + 2 s^2 t u (t + u)^2 (8 t^6 + 65 t^5 u + 167 t^4 u^2\nonumber\\
 & & + 193 t^3 u^3 + 111 t^2 u^4 + 29 t u^5 + 2 u^6)
     + s t^2 u^2 (t + u)^4 (11 t^3 + 34 t^2 u + 28 t u^2 + 7 u^3)\nonumber\\
 & & +2 t^3 u^3 (t + u)^6],
\nonumber\\
L&=& 8 Q^{10} t^4 (s + t) (t + u) [3 s^3+ 
     s^2 (7 t + 10 u) + s (5 t^2 + 15 t u + 9 u^2) 
     + t^3 + 5 t^2 u + 7 t u^2 + 2 u^3]\nonumber\\
 & & + 2 Q^8 t^2 [s^8 - 2 s^7 (t + u) - s^6 (21 t^2 + 42 t u + 23 u^2)- 
     2 s^5 (29 t^3 + 62 t^2 u + 55 t u^2 + 22 u^3)\nonumber\\
 & & - s^4 (98 t^4 + 218 t^3 u + 176 t^2 u^2 + 90 t u^3 + 33 u^4) - 
     2 s^3 (54 t^5 + 157 t^4 u + 130 t^3 u^2 + 35 t^2 u^3\nonumber\\
 & & + 13 t u^4 + 5 u^5) - 
     s^2 (t + u)^2 (70 t^4 + 156 t^3 u + 4 t^2 u^2 + 6 t u^3 + u^4)- 
     2 s t (11 t^6 + 69 t^5 u\nonumber\\
 & & + 144 t^4 u^2 + 123 t^3 u^3 + 42 t^2 u^4 + 
     6 t u^5 + u^6)-2 t^3 (t + u)^2 (t^3 + 9 t^2 u + 17 t u^2 + 7 u^3)]
     \nonumber\\
 & & - 2 Q^6 t [s^9 (3 t - u) + s^8 (t^2 + 3 t u + 2 u^2) - 
     s^7 (35 t^3 + 47 t^2 u - 3 t u^2 - 23 u^3) - 
     s^6 (101 t^4\nonumber\\
 & & + 233 t^3 u + 167 t^2 u^2 - 9 t u^3 - 44 u^4) - s^5 (134 t^5 
     + 422 t^4 u + 528 t^3 u^2 + 294 t^2 u^3 + 17 t u^4\nonumber\\
 & & - 33 u^5)- s^4 (96 t^6 + 392 t^5 u + 648 t^4 u^2 + 
     620 t^3 u^3 + 313 t^2 u^4 + 35 t u^5 - 10 u^6)- s^3 (t + u)^2\nonumber\\
 & & (30 t^5 + 126 t^4 u + 114 t^3 u^2 +142 t^2 u^3 + 15 t u^4 - u^5)+ 
     s^2 t (6 t^7 - 22 t^6 u - 130 t^5 u^2 - 228 t^4 u^3\nonumber\\
 & & - 241 t^3 u^4 - 
     161 t^2 u^5 - 47 t u^6 - u^7)+ 2 s t^2 (t + u)^2 (4 t^5 + 2 t^4 u 
     - 17 t^3 u^2 - 19 t^2 u^3 - 
     12 t u^4\nonumber\\
 & & - 4 u^5)+2 t^3 (t + u)^4 (t^3 - 4 t u^2 - u^3)]+ 
     Q^4 [s^{10} (5 t^2 - 8 t u - u^2)+ 
     2 s^9 (4 t^3 - 6 t^2 u - 15 t u^2\nonumber\\
 & & - 5 u^3)-s^8 (32 t^4 + 48 t^3 u + 59 t^2 u^2 + 54 t u^3 + 35 u^4)- 
     4 s^7 (30 t^5 + 63 t^4 u + 45 t^3 u^2 + 10 t^2 u^3\nonumber\\
 & & + 11 t u^4 + 13 u^5)
     - s^6 (179 t^6 + 596 t^5 u + 694 t^4 u^2 + 228 t^3 u^3 - 
     120 t^2 u^4 - 24 t u^5 + 35 u^6)\nonumber\\
 & & - 2 s^5 (78 t^7 + 404 t^6 u + 756 t^5 u^2 + 600 t^4 u^3 + 74 t^3 u^4 - 
     132 t^2 u^5 - 31 t u^6 + 5 u^7)- s^4 (t + u)^2\nonumber\\ 
 & & (86 t^6 + 516 t^5 u + 768 t^4 u^2 + 360 t^3 u^3 - 142 t^2 u^4 - 
     32 t u^5 + u^6)- 
     4 s^3 t (7 t^8 + 89 t^7 u + 338 t^6 u^2\nonumber\\
 & & + 596 t^5 u^3 + 535 t^4 u^4 + 
     212 t^3 u^5 - t^2 u^6 - 19 t u^7 - u^8)- 
     s^2 t^2 (t + u)^2 (4 t^6 + 92 t^5 u + 338 t^4 u^2\nonumber\\
 & & + 444 t^3 u^3 + 
     245 t^2 u^4 + 30 t u^5 - 13 u^6)- 
     4 s t^3 u (t + u)^4 (3 t^3 + 12 t^2 u + 8 t u^2 + 2 u^3)
     -4 t^4 u^2 (t\nonumber\\
 & & + u)^6]+ 2 Q^2 s (s + t + u)^2 [s^8 u (5 t + u)
     + s^7 (2 t^3 + 16 t^2 u + 21 t u^2 + 7 u^3)+ 
     s^6 (5 t^4 + 24 t^3 u\nonumber\\
 & & + 57 t^2 u^2 + 46 t u^3 + 16 u^4) + 
     s^5 (3 t^5 + 13 t^4 u + 55 t^3 u^2 + 79 t^2 u^3 + 50 t u^4 + 16 u^5) - 
     s^4 (t^6 + 11 t^5 u\nonumber\\
 & & + 22 t^4 u^2 + 4 t^3 u^3 - 19 t^2 u^4 - 22 t u^5 - 
     7 u^6) - s^3 (t^7 + 15 t^6 u + 70 t^5 u^2 + 128 t^4 u^3 + 113 t^3
     u^4\nonumber\\
 & & + 37 t^2 u^5 - 5 t u^6 - u^7)- 
     s^2 t u (t + u)^2 (4 t^4 + 27 t^3 u + 49 t^2 u^2 + 26 t u^3 - u^4)
     - s t^2 u^2 (t + u)^2\nonumber\\
 & & (5 t^3 + 19 t^2 u + 18 t u^2 + 
     6 u^3)-2 t^3 u^3 (t + u)^4] - s^2 (s + u) (s + t + u)^2 
     [s^7 (t^2 + 4 t u + u^2)\nonumber\\
 & & + s^6 (4 t^3 + 19 t^2 u + 18 t u^2 + 5 u^3)+ 
     2 s^5 (3 t^4 + 19 t^3 u + 33 t^2 u^2 + 21 t u^3 + 5 u^4)+ 
     2 s^4 (2 t^5\nonumber\\
 & & + 20 t^4 u + 56 t^3 u^2 + 58 t^2 u^3 + 26 t u^4 + 5 u^5)+ 
     s^3 (t^6 + 22 t^5 u + 98 t^4 u^2 + 156 t^3 u^3 + 116 t^2 u^4\nonumber\\
 & & + 42 t u^5 + 5 u^6)+s^2 u (t + u)^2 (5 t^4 + 30 t^3 u + 33 t^2 u^2 
     + 16 t u^3 + u^4)+s t u^2 (t + u)^3 (5 t^2\nonumber\\
 & & + 7 t u + 4 u^2)+t^2 u^3 (t + u)^4],
\nonumber\\
A&=&0.
\end{eqnarray}

$\nu(\overline{\nu})+g\to\nu(\overline{\nu})
+c\overline{c}\left[{}^3\!P_2^{(1)}\right]+g$:
\begin{eqnarray}
F&=& \frac{4 g^4 \alpha_s^2 a_c^2}{135 \pi M^3 (Q^2 + M_Z^2)^2 s^4 (s + t)^4 
     (s + u)^4 (t + u)^4},
\nonumber\\
T&=& -24 Q^{10} t^4 (s + t) (t + u) [s^3+ 
     s^2 (3 t + 4 u) + s (3 t^2 + 7 t u + 5 u^2)+ t^3 + 3 t^2 u + 3 t u^2 
     + 2 u^3]\nonumber\\
 & & +4 Q^8 t^2 [s^8 + 5 s^7 (t + u) + 2 s^6 (8 t^2 + 16 t u + 7 u^2) + 
     s^5 (28 t^3 + 84 t^2 u + 82 t u^2 + 26 u^3)\nonumber\\
 & & + s^4 (61 t^4 + 168 t^3 u + 181 t^2 u^2 + 102 t u^3 + 29 u^4) + 
     s^3 (112 t^5 + 365 t^4 u + 443 t^3 u^2 + 241 t^2 u^3\nonumber\\
 & & + 68 t u^4 + 17 u^5)+ 
     s^2 (t + u)^2 (103 t^4 + 234 t^3 u + 146 t^2 u^2 + 20 t u^3 + 4 u^4) 
     + s t (43 t^6 + 229 t^5 u\nonumber\\
 & & + 498 t^4 u^2 + 551 t^3 u^3 + 
     298 t^2 u^4 + 66 t u^5 + 7 u^6)+ 
     t^2 (t + u)^2 (7 t^4 + 29 t^3 u + 47 t^2 u^2 + 47 t u^3\nonumber\\
 & & + 4 u^4)] - 
     2 Q^6 t [s^9 (4 t - 2 u) + 2 s^8 (8 t^2 + 3 t u - 5 u^2) + 
     s^7 (61 t^3 + 94 t^2 u - 7 t u^2 - 28 u^3)\nonumber\\
 & & + s^6 (151 t^4 + 443 t^3 u + 319 t^2 u^2 - 25 t u^3 - 52 u^4)+ 
     s^5 (225 t^5 + 842 t^4 u + 1102 t^3 u^2 + 518 t^2 u^3\nonumber\\
 & & - 19 t u^4 -58 u^5)+ 
     s^4 (247 t^6 + 1025 t^5 u + 1648 t^4 u^2 + 1210 t^3 u^3 + 385 t^2 u^4 + 
     11 t u^5 - 34 u^6)\nonumber\\
 & & + s^3 (t + u)^2 (180 t^5 + 628 t^4 u + 545 t^3 u^2 + 106 t^2 u^3 
     + 38 t u^4 - 8 u^5) 
     + s^2 t (50 t^7 + 548 t^6 u\nonumber\\
 & & + 1725 t^5 u^2 + 
     2371 t^4 u^3 + 1543 t^3 u^4 + 483 t^2 u^5 + 92 t u^6 + 8 u^7)
     - 2 s t^2 (t + u)^2 (6 t^5 - 56 t^4 u\nonumber\\
 & & - 229 t^3 u^2 -255 t^2 u^3 - 68 t u^4 - 16 u^5)- 
     2 t^3 (t + u)^3 (3 t^4 - 3 t^3 u - 32 t^2 u^2 - 49 t u^3 - 11
     u^4)]\nonumber\\
 & & - 2 Q^4 [s^{10} (t^2 + 10 t u + 3 u^2) + 
     s^9 (9 t^3 + 55 t^2 u + 61 t u^2 + 15 u^3)+ 
     3 s^8 (4 t^4 + 48 t^3 u + 99 t^2 u^2\nonumber\\
 & & + 62 t u^3 + 11 u^4) - 
     s^7 (64 t^5 - 27 t^4 u - 581 t^3 u^2 - 803 t^2 u^3 - 355 t u^4 - 
     42 u^5)-s^6 (221 t^6\nonumber\\
 & & + 557 t^5 u - 54 t^4 u^2 - 1270 t^3 u^3 - 1278 t^2 u^4 - 
     437 t u^5 - 33 u^6) - s^5 (317 t^7 + 1180 t^6 u\nonumber\\
 & & + 1263 t^5 u^2 - 400 t^4 u^3 - 1694 t^3 u^4 - 1215 t^2 u^5 - 336 t u^6 
     - 15 u^7)- 
     s^4 (t + u)^2 (266 t^6 + 745 t^5 u\nonumber\\
 & & + 340 t^4 u^2 - 464 t^3 u^3 - 
     408 t^2 u^4 - 141 t u^5 - 3 u^6)- 
     s^3 t (128 t^8 + 840 t^7 u + 1933 t^6 u^2 + 1756 t^5 u^3\nonumber\\
 & & + 69 t^4 u^4 - 
     969 t^3 u^5 - 694 t^2 u^6 - 219 t u^7 - 28 u^8)
     - s^2 t^2 (t + u)^2 (26 t^6 + 240 t^5 u + 540 t^4 u^2\nonumber\\
 & & + 321 t^3 u^3 - 111 t^2 u^4 - 81 t u^5 - 29 u^6)- 2 s t^3 u (t + u)^3 
     (17 t^4 + 73 t^3 u + 76 t^2 u^2 + 10 t u^3 + 2 u^4)\nonumber\\
 & & -14 t^4 u^2 (t + u)^6] + 
     2 Q^2 s [s^{10} (3 t^2 + 8 t u + 3 u^2) + 
     6 s^9 (3 t^3 + 10 t^2 u + 10 t u^2 + 3 u^3)
     + s^8 (54 t^4\nonumber\\
 & & + 228 t^3 u + 346 t^2 u^2 + 216 t u^3 + 48 u^4) + 
     s^7 (93 t^5 + 526 t^4 u + 1105 t^3 u^2 + 1061 t^2 u^3 + 464 t
     u^4\nonumber\\
 & & + 75 u^5)+ 
     s^6 (99 t^6 + 764 t^5 u + 2132 t^4 u^2 + 2872 t^3 u^3 + 1980 t^2 u^4 + 
     652 t u^5 + 75 u^6) + s^5 (72 t^7\nonumber\\
 & & + 734 t^6 u + 2618 t^5 u^2 + 
     4630 t^4 u^3 + 4480 t^3 u^4 + 2370 t^2 u^5 + 612 t u^6 + 48 u^7)+ 
     s^4 (t + u)^2 (36 t^6\nonumber\\
 & & + 404 t^5 u + 1297 t^4 u^2 + 1774 t^3 u^3 + 
     1143 t^2 u^4 + 332 t u^5 + 18 u^6)+ 
     s^3 (9 t^9 + 188 t^8 u + 1142 t^7 u^2\nonumber\\
 & & + 3251 t^6 u^3 + 5142 t^5 u^4 + 
     4834 t^4 u^5 + 2729 t^3 u^6 + 874 t^2 u^7 + 128 t u^8 + 3 u^9)
     + 2 s^2 t u (t + u)^2 (16 t^6\nonumber\\
 & & + 138 t^5 u + 387 t^4 u^2 + 
     495 t^3 u^3 + 311 t^2 u^4 + 100 t u^5 + 10 u^6)+ 
     s t^2 u^2 (t + u)^3 (37 t^4 + 161 t^3 u\nonumber\\
 & & + 234 t^2 u^2 + 127 t u^3 + 
     29 u^4)+14 t^3 u^3 (t + u)^6],
\nonumber\\
L&=& 24 Q^{10} t^4 (s + t) (t + u) [s^3+s^2 (3 t + 4 u) + s (3 t^2 + 7 t u 
     + 5 u^2)
     + t^3 + 3 t^2 u + 3 t u^2 + 2 u^3]\nonumber\\
 & & - 2 Q^8 t^2 [s^8 + 14 s^7 (t + u) + s^6 (55 t^2 + 110 t u + 53 u^2) + 
     2 s^5 (67 t^3 + 162 t^2 u + 141 t u^2 + 46 u^3)\nonumber\\
 & & + s^4 (254 t^4 + 666 t^3 u + 652 t^2 u^2 + 322 t u^3 + 83 u^4)+ 
     2 s^3 (160 t^5 + 521 t^4 u + 608 t^3 u^2 + 319 t^2 u^3\nonumber\\
 & & + 91 t u^4 + 
     19 u^5)+ s^2 (t + u)^2 (230 t^4 + 516 t^3 u + 304 t^2 u^2 + 46 t u^3 
     + 7 u^4)+ 2 s t (43 t^6 + 229 t^5 u\nonumber\\
 & & + 498 t^4 u^2 + 551 t^3 u^3 + 
     298 t^2 u^4 + 66 t u^5 + 7 u^6)+ 
     2 t^2 (t + u)^2 (7 t^4 + 29 t^3 u + 47 t^2 u^2 + 47 t u^3\nonumber\\
 & & + 4 u^4)]
     + 2 Q^6 t [s^9 (3 t - u) + s^8 (51 t^2 + 37 t u - 14 u^2) + 
     s^7 (221 t^3 + 389 t^2 u + 107 t u^2 - 53 u^3)\nonumber\\
 & & + s^6 (467 t^4 + 1269 t^3 u + 1049 t^2 u^2 + 155 t u^3 - 92 u^4)+ 
     s^5 (610 t^5 + 2090 t^4 u + 2674 t^3 u^2\nonumber\\
 & & + 1434 t^2 u^3 + 161 t u^4 - 83 u^5) + 
     s^4 (542 t^6 + 2196 t^5 u + 3484 t^4 u^2 + 2782 t^3 u^3 + 1115 t^2
     u^4\nonumber\\
 & & + 125 t u^5 - 38 u^6)+ s^3 (t + u)^2 
     (302 t^5 + 990 t^4 u + 870 t^3 u^2 + 342 t^2 u^3 + 67 t u^4 - 7 u^5)
     + s^2 t (70 t^7\nonumber\\
 & & + 670 t^6 u + 2020 t^5 u^2 + 
     2756 t^4 u^3 + 1859 t^3 u^4 + 643 t^2 u^5 + 127 t u^6 + 7 u^7)
     - 2 s t^2 (t + u)^2 (6 t^5\nonumber\\
 & & - 56 t^4 u - 229 t^3 u^2 - 255 t^2 u^3 - 68 t u^4 - 16 u^5)- 
     2 t^3 (t + u)^3 (3 t^4 - 3 t^3 u - 32 t^2 u^2 - 49 t u^3\nonumber\\
 & & - 11 u^4)]- 
     Q^4 [s^{10} (13 t^2 + 8 t u + 7 u^2) + 
     2 s^9 (74 t^3 + 56 t^2 u + t u^2 + 19 u^3)+ 
     s^8 (588 t^4 + 880 t^3 u\nonumber\\
 & & + 61 t^2 u^2 - 118 t u^3 + 89 u^4) + 
     4 s^7 (298 t^5 + 750 t^4 u + 413 t^3 u^2 - 161 t^2 u^3 - 93 t u^4 + 
     29 u^5)\nonumber\\
 & & + s^6 (1421 t^6 + 5216 t^5 u + 5998 t^4 u^2 + 1312 t^3 u^3 - 1504 t^2 u^4 
     - 536 t u^5 + 89 u^6) + 2 s^5 (536 t^7\nonumber\\
 & & + 2616 t^6 u + 4606 t^5 u^2 + 
     3264 t^4 u^3 + 232 t^3 u^4 - 734 t^2 u^5 - 209 t u^6 + 19 u^7)+ 
     s^4 (t + u)^2 (534 t^6\nonumber\\
 & & + 2328 t^5 u + 2692 t^4 u^2 + 896 t^3 u^3 - 
     370 t^2 u^4 - 184 t u^5 + 7 u^6)+ 
     4 s^3 t (43 t^8 + 397 t^7 u + 1177 t^6 u^2\nonumber\\
 & & + 1596 t^5 u^3 + 1060 t^4 u^4 + 
     279 t^3 u^5 - 55 t^2 u^6 - 54 t u^7 - 7 u^8)+ s^2 t^2 (t + u)^2 (28 t^6 
     + 444 t^5 u\nonumber\\
 & & + 1178 t^4 u^2 + 
     920 t^3 u^3 + 103 t^2 u^4 - 26 t u^5 - 43 u^6)+ 
     4 s t^3 u (t + u)^3 (17 t^4 + 73 t^3 u + 76 t^2 u^2\nonumber\\
 & & + 10 t u^3 + 2 u^4)+28 t^4 u^2 (t + u)^6]+ 
     2 Q^2 s [s^{10} (8 t^2 + 11 t u + 7 u^2)+ 
     s^9 (58 t^3 + 78 t^2 u + 65 t u^2\nonumber\\
 & & + 45 u^3)+ 
     s^8 (177 t^4 + 287 t^3 u + 162 t^2 u^2 + 187 t u^3 + 127 u^4) + 
     s^7 (295 t^5 + 655 t^4 u + 232 t^3 u^2\nonumber\\
 & & - 20 t^2 u^3 + 313 t u^4 + 205 u^5)+ 
     s^6 (290 t^6 + 925 t^5 u + 438 t^4 u^2 - 844 t^3 u^3 - 540 t^2 u^4 + 
     308 t u^5\nonumber\\
 & & + 205 u^6)+ 
     s^5 (168 t^7 + 776 t^6 u + 692 t^5 u^2 - 1164 t^4 u^3 - 2216 t^3 u^4 - 
     930 t^2 u^5 + 165 t u^6 + 127 u^7)\nonumber\\
 & & + s^4 (t + u)^2 (53 t^6 + 247 t^5 u 
     - 61 t^4 u^2 - 
     871 t^3 u^3 - 719 t^2 u^4 - 55 t u^5 + 45 u^6)+ s^3 (t + u)^2 (7
     t^7\nonumber\\ 
 & & + 53 t^6 u - 73 t^5 u^2 - 
     531 t^4 u^3 - 697 t^3 u^4 - 339 t^2 u^5 - 17 t u^6 + 7 u^7)- 
     s^2 t u^2 (t + u)^3 (85 t^4 + 274 t^3 u\nonumber\\
 & & + 227 t^2 u^2 + 99 t u^3 + u^4)- 
     3 s t^2 u^2 (t + u)^4 (7 t^3 + 33 t^2 u + 22 t u^2 + 6 u^3)
     -14 t^3 u^3 (t + u)^6]\nonumber\\
 & & - s^2 (s + u) (s + t + u)^2 
     [s^7 (7 t^2 + 12 t u + 7 u^2) + s^6 (28 t^3 + 45 t^2 u + 50 t u^2 + 
     31 u^3)+ 2 s^5 (21 t^4\nonumber\\
 & & + 33 t^3 u + 51 t^2 u^2 + 69 t u^3 + 29 u^4) + 
     2 s^4 (14 t^5 + 24 t^4 u + 34 t^3 u^2 + 90 t^2 u^3 + 94 t u^4 + 29
     u^5)\nonumber\\
 & & + s^3 (7 t^6 + 18 t^5 u - 2 t^4 u^2 + 60 t^3 u^3 + 180 t^2 u^4 
     + 138 t u^5 + 
     31 u^6)+ s^2 u (t + u)^2 (3 t^4 - 14 t^3 u\nonumber\\
 & & + 23 t^2 u^2 + 36 t u^3 + 
     7 u^4)+ 3 s t u^2 (t + u)^3 (t^2 + 3 t u + 4 u^2)+7 t^2 u^3 (t + u)^4],
\nonumber\\
A&=&0.
\end{eqnarray}

$\nu(\overline{\nu})+g\to\nu(\overline{\nu})
+c\overline{c}\left[{}^1\!S_0^{(8)}\right]+g$:
\begin{eqnarray}
F&=&\frac{g^4 \alpha_s^2}{12\pi M 
(Q^2 + M_Z^2)^2 s^4 t (s + t)^2 (s + u)^2 (t + u)^2},
\nonumber\\
T&=& 9 v_c^2 Q^2 (Q^2 t + s u) [ 2 Q^4 t^2 u^2 
     + 2 Q^2 s t u (s t+s u+t^2+t u+u^2) 
     + s^6 
\nonumber\\
 & & {}+ 2 s^5 (t + u) + 3 s^4 (t + u)^2 + 
     2 s^3 (t + u)^3  + 
     s^2 (t^4 + 2 t^3 u + 3 t^2 u^2
     + 2 t u^3 + u^4)]
\nonumber\\
 & & {}+10 a_c^2 Q^2 s^2 t^2 (Q^2 t + s u) (s + t + u)^2,
\nonumber\\
L&=& - 9 v_c^2 Q^4 t [ 2 Q^4 t^2 u^2 
+ 2 Q^2 s t u (s t+s u+t^2 +t u +2 u^2)+ s^4 (t + u)^2
\nonumber\\
& & {} + 2 s^3 (t^3 + 2 t^2 u + 2 t u^2 + u^3)
     + s^2 (t^4 + 2 t^3 u + 3 t^2 u^2
 + 2 t u^3 + 2 u^4)]
\nonumber\\
 & & {} -5 a_c^2 s^2 t (Q^2 t + s u)^2 (s + t + u)^2,
\nonumber\\
A&=&0.
\end{eqnarray}

$\nu(\overline{\nu})+g\to\nu(\overline{\nu})
+c\overline{c}\left[{}^3\!S_1^{(8)}\right]+g$:
\begin{eqnarray}
F&=&\frac{-g^4 \alpha_s^2}{36\pi M (Q^2 + M_Z^2)^2
s^4 t (s + t)^2 (s + u)^2 (t + u)^2},
\nonumber\\
T&=& 10 v_c^2 Q^2 t \{
     2 Q^6 t^2 (s^2 + t^2) 
     - Q^4 t [s^3 (3 t - 2 u) + 3 s^2 t (t + u) + 2 s t^2 (t - u)
     + 2 t^3 (t + u)]
\nonumber\\ 
 & & {}+ Q^2 s [s^3 (t - u)^2 - 2 s^2 t u (t + u) - s t^2 u (2 t - u)
     - 2 t^3 u (t + u)] 
     -s^5 (t^2 + t u + u^2)
\nonumber\\
 & & {}- s^4 (t + u)^3 - s^3 t u (t^2 + 3 t u + u^2)
     - s^2 t^2 u^2 (t + u) 
     \} 
     + 9 a_c^2 Q^2 \{
     2 Q^6 t^3 u^2 - 2 Q^4 t^2 u [s^3 
\nonumber\\
 & & {} - s^2 (t + u) + s (t^2 + t u - 2 u^2) + 2 t u (t + u) 
     ] 
     - Q^2 s t [s^5 + 2 s^4 (t + u) + s^3 (t^2 + 2 t u
\nonumber\\
 & & {}+ 3 u^2) + 2 s^2 t (t^2 + 3 t u + 2 u^2) + s (t^4 + 6 t^3 u 
     + 15 t^2 u^2 + 12 t u^3 
     - u^4) + 4 t u^2 (t^2 + 3 t u 
\nonumber\\ 
 & & {} + 2 u^2)] -s^7 u - 2 s^6 u (t + u) - s^5 (2 t^3 + 
     5 t^2 u + 6 t u^2 + 3 u^3)
     - 2 s^4 (t^4 + 4 t^3 u + 7 t^2 u^2 
\nonumber\\
 & & {}+ 5 t u^3 + u^4)
     - s^3 u (3 t^4 + 12 t^3 u + 17 t^2 u^2 + 10 t u^3
     + u^4) - 2 s^2 t u^2 (t^3 + 3 t^2 u + 4 t u^2
\nonumber\\
 & & {} + 2 u^3)\},
\nonumber\\
L&=& 5 v_c^2 Q^4 t \{
     2 Q^4 t^2 (s^2 - 2 t^2) 
     - 2 Q^2 t (s^2 - 2 t^2) [s(t - u) + t(t+ u)] 
     + s^4 (t^2 + u^2)
\nonumber\\
 & & {}+ 2 s^3 t^2 (t + u) + s^2 t^2 (t^2 + 6 t u + u^2)
     + 4 s t^3 u (t + u)
     \} 
     - 9 a_c^2 \{
     2 Q^8 t^3 u^2 
     + 2 Q^6 t^2 u [s^3
\nonumber\\
 & & {}+ s^2 (t + u) - s (t^2 + t u - 2 u^2)
     - 2 t u (t + u)]
     + Q^4 s t [s^5 + 2 s^4 (t
     + u) + s^3 u (t + 5 u)
\nonumber\\ 
 & & {}- s^2 (2 t^3 + 7 t^2 u + t u^2 - 4 u^3)
     - s (t^4 + 6 t^3 u + 14 t^2 u^2 + 8 t u^3 - 3 u^4)
     - 4 t u^2 (t^2 + 3 t u
\nonumber\\
 & & {}+ 2 u^2)]
     - Q^2 s^2 [s^5 (t - u)
     + s^4 (3 t^2 + t u - 2 u^2)
     + s^3 (3 t^3 + 8 t^2 u + 4 t u^2 - 3 u^3) + s^2 (t^4
\nonumber\\
 & & {}+ 9 t^3 u + 18 t^2 u^2 + 8 t u^3
     - 2 u^4)
     + s u (3 t^4 + 13 t^3 u + 20 t^2 u^2 + 9 t u^3 - u^4)
     + t u^2 (t 
\nonumber\\
 & & {}+ u)^2 (t+ 5 u)]
     - s^3 u (s + t
     + u) [s^2 + s (t + u) + u (t + u)]^2 
     \},
\nonumber\\
A&=&0.
\end{eqnarray}

$\nu(\overline{\nu})+g\to\nu(\overline{\nu})
+c\overline{c}\left[{}^3\!P_J^{(8)}\right]+g$:
\begin{eqnarray}
F&=&\frac{-g^4 \alpha_s^2}{3\pi M^3
(Q^2 + M_Z^2)^2 s^4 t (s + t)^3 (s + u)^4 (t + u)^3},
\nonumber\\
T&=& 9 v_c^2 Q^2 \{
     8 Q^8 t^2 [2 s^5 t - 2 s^4 u^2 + s^3 t^2 (2 t - 3 u)
     - s^2 t u (t^2 + 3 t u - 3 u^2)
     - s t u (2 t^3 + t^2 u 
\nonumber\\
 & & {}- t u^2 - u^3)
     + t^2 u^3 (2 t + u)] 
     - 2 Q^6 t [4 s^7 (t + u) + 4 s^6 (5 t^2 + 2 u^2) + 
     4 s^5 (5 t^3 + 6 t^2 u 
\nonumber\\ 
 & & {}- 2 t u^2 + 4 u^3) + 2 s^4 (9 t^4 - 9 t^3 u + 4 t^2 u^2 
     - 6 t u^3 
     + 4 u^4) + s^3 (12 t^5 - 2 t^4 u - 29 t^3 u^2 
\nonumber\\
 & & {} + 33 t^2 u^3 - 12 t u^4 + 4 u^5) - s^2 t (2 t^5 + 20 t^4 u 
     + 25 t^3 u^2 - t^2 u^3 - 16 t u^4 + 4 u^5)
     - s t^2 u 
\nonumber\\
 & & {} \times (12 t^4 + 26 t^3 u + 2 t^2 u^2 + t u^3 + u^4)
     -t^3 u^2 (2 t^3 - 6 t^2 u - 15 t u^2 - 7 u^3)] 
     + 2 Q^4 [2 s^8 
\nonumber\\
 & & {} \times (3 t^2 + t u - 2 u^2)+ s^7 (21 t^3 + 3 t^2 u 
     + 10 t u^2 - 8 u^3)
     + s^6 (28 t^4 + 15 t^3 u - 15 t^2 u^2 + 22 t u^3
\nonumber\\
 & & {} - 12 u^4) + s^5 (27 t^5 + 11 t^4 u + 8 t^3 u^2 
     - 20 t^2 u^3 + 26 t u^4 - 8 u^5)
     + s^4 (16 t^6 + 17 t^5 u - 20 t^4 u^2 
\nonumber\\
 & & {}+ 2 t^3 u^3 -33 t^2 u^4 + 18 t u^5 - 4 u^6)
     + s^3 t (2 t^6 + 8 t^5 u - 3 t^4 u^2 - 19 t^3 u^3 - 19 t^2 u^4 - 
     17 t u^5 
\nonumber\\    
 & & {} + 6 u^6)- s^2 t^2 (2 t^6 + 14 t^5 u + 
     30 t^4 u^2 + 33 t^3 u^3 + 46 t^2 u^4 + 21 t u^5 + 4 u^6)    
     - 2 s t^3 u (2 t^5 
\nonumber\\
 & & {}+ 10 t^4 u + 16 t^3 u^2 + 20 t^2 u^3 + 19 t u^4 + 
     7 u^5)    
     -2 t^6 u^2 (t + u)^2]
     - Q^2 s [s^8 (7 t^2 - 5 t u - 12 u^2)
\nonumber\\
 & & {}+ s^7 (25 t^3 + 3 t^2 u - 18 t u^2 - 
     36 u^3)+ s^6 (37 t^4 + 25 t^3 u - 12 t^2 u^2 - 20 t u^3 - 60 u^4)
     +s^5 (39 t^5
\nonumber\\
 & & {}+ 39 t^4 u + 16 t^3 u^2 - 18 t^2 u^3 - 6 t u^4 - 60 u^5)
     + s^4 (29 t^6 + 83 t^5 u + 72 t^4 u^2 + 88 t^3 u^3 + 
     40 t^2 u^4
\nonumber\\
 & & {}+ 22 t u^5 - 36 u^6)
     +s^3 (9 t^7 + 75 t^6 u + 148 t^5 u^2 + 176 t^4 u^3 + 178 t^3 u^4 + 
     102 t^2 u^5 + 22 t u^6 
\nonumber\\
 & & {}- 12 u^7)
     + s^2 t u (22 t^6 + 107 t^5 u + 
     199 t^4 u^2 + 211 t^3 u^3 + 177 t^2 u^4 + 73 t u^5 + 9 u^6)    
     + s t^2 u^2 (17 t^5
\nonumber\\
 & & {}+ 69 t^4 u + 107 t^3 u^2 + 105 t^2 u^3 + 71 t u^4 + 
     21 u^5)    
     + 4 t^5 u^3 (t + u)^2]       
     - s^2 (s + u) [7 s^7 u (t + u)
\nonumber\\ 
 & & {}+ s^6 u (25 t^2 + 38 t u + 21 u^2)
     + s^5 (2 t^4 + 47 t^3 u + 88 t^2 u^2 + 78 t u^3 + 35 u^4)
     + s^4 (4 t^5 + 63 t^4 u 
\nonumber\\
 & & {}+ 132 t^3 u^2 + 156 t^2 u^3 + 
     98 t u^4 + 35 u^5)   
     + s^3 (2 t^6 + 47 t^5 u + 136 t^4 u^2 + 190 t^3 u^3 + 156 t^2 u^4 
\nonumber\\
 & & {}+ 78 t u^5 + 21 u^6)   
     + s^2 u (13 t^6 + 70 t^5 u + 136 t^4 u^2 + 
     132 t^3 u^3 + 88 t^2 u^4 + 38 t u^5 + 7 u^6)   
     + s t u^2 
\nonumber\\
 & & {} \times (13 t^5 + 47 t^4 u + 63 t^3 u^2 + 47 t^2 u^3 
     + 25 t u^4 + 7 u^5)   
     +2 t^4 u^3 (t + u)^2]\} 
     + 10 a_c^2 Q^2 t \{
     4 Q^8 t^4
\nonumber\\
 & & {} \times [2 s^3 + s^2 (5 t + 7 u) + s (4 t^2 + 11 t u + 7 u^2)
     + (t + u)^2 (t + 2 u)] 
     + 2 Q^6 t^2 [2 s^6 + s^5 (7 t 
\nonumber\\
 & & {}+ 9 u) + s^4 (5 t^2 + 27 t u + 12 u^2) - s^3 (13 t^3 
     - 7 t^2 u 
     - 34 t u^2 - 
     4 u^3) - s^2 (23 t^4 + 57 t^3 u 
\nonumber\\
 & & {} + 18 t^2 u^2 - 14 t u^3 + 2 u^4) - 
     s (12 t^5 + 52 t^4 u + 73 t^3 u^2 + 33 t^2 u^3 + t u^4 + u^5) - 
     t (2 t^5 
\nonumber\\
 & & {}+ 12 t^4 u + 27 t^3 u^2+ 29 t^2 u^3 + 13 t u^4 + u^5)] 
     - Q^4 t [2 s^7 (5 t - 2 u) + s^6 (41 t^2 + 37 t u - 18 u^2)
\nonumber\\
 & & {}+ s^5 (56 t^3 + 147 t^2 u+ 45 t u^2 - 24 u^3) + 
     s^4 (19 t^4 + 164 t^3 u + 194 t^2 u^2 + 19 t u^3 - 8 u^4) 
\nonumber\\
 & & {}- s^3 (16 t^5 - 12 t^4 u - 154 t^3 u^2 - 117 t^2 u^3 
     + 5 t u^4 - 4 u^5)
     - s^2 (8 t^6 + 66 t^5 u + 95 t^4 u^2 
\nonumber\\
 & & {}+ 8 t^3 u^3 - 
     17 t^2 u^4 + 10 t u^5 - 2 u^6) +2 s t (2 t^6 - 8 t^5 u - 53 t^4 u^2 
     - 82 t^3 u^3 - 45 t^2 u^4
\nonumber\\
 & & {}- 8 t u^5 - 2 u^6)
     +2 t^2 (t + u)^2 (t^4 - 8 t^2 u^2 - 12 t u^3 - 3 u^4)] 
     + Q^2 [s^9 (t + u) + s^8 (11 t^2
\nonumber\\
 & & {}+ t u + 6 u^2) 
     + s^7 (40 t^3 + 36 t^2 u - 3 t u^2 + 17 u^3)
     + s^6 (56 t^4 + 144 t^3 u + 57 t^2 u^2 + 9 t u^3
\nonumber\\
 & & {}+ 24 u^4)
     + s^5 (19 t^5 + 167 t^4 u + 230 t^3 u^2 + 90 t^2 u^3 + 41 t u^4 + 
     17 u^5) 
     - s^4 (29 t^6 + 5 t^5 u
\nonumber\\
 & & {}- 182 t^4 u^2 - 228 t^3 u^3 - 
     117 t^2 u^4 - 53 t u^5 - 6 u^6)
     - s^3 (30 t^7 + 126 t^6 u + 
     115 t^5 u^2 
\nonumber\\
 & & {}- 81 t^4 u^3 - 158 t^3 u^4 - 88 t^2 u^5 - 31 t u^6 - u^7)
     - s^2 t (8 t^7 + 72 t^6 u + 191 t^5 u^2 + 181 t^4 u^3
\nonumber\\
 & & {}+ 16 t^3 u^4 - 
     60 t^2 u^5 - 29 t u^6 - 7 u^7)
     - 2 s t^2 u (t + u)^2 
     (5 t^4 + 21 t^3 u + 21 t^2 u^2 + 2 t u^3
\nonumber\\
 & & {}- u^4)
     - 4 t^3 u^2 (t + u)^5]
     - s (s + u) [s^8 (t + u) + 6 s^7 (t^2 + t u + u^2)
     + s^6 (19 t^3 + 29 t^2 u
\nonumber\\
 & & {}+ 27 t u^2 + 17 u^3)
     + 2 s^5 (17 t^4 + 47 t^3 u + 50 t^2 u^2 + 35 t u^3 + 12 u^4)
     + s^4 (35 t^5 + 162 t^4 u
\nonumber\\
 & & {}+ 254 t^3 u^2 + 202 t^2 u^3 + 92 t u^4 + 
     17 u^5)
     + s^3 (20 t^6 + 145 t^5 u + 338 t^4 u^2 + 370 t^3 u^3
\nonumber\\ 
 & & {}+ 216 t^2 u^4 + 65 t u^5 + 6 u^6)
     + s^2 (5 t^7 + 62 t^6 u + 217 t^5 u^2 + 346 t^4 u^3 + 286 t^3 u^4 + 
     125 t^2 u^5
\nonumber\\
 & & {}+ 26 t u^6 + u^7)
     + s t u (t + u)^2 (9 t^4 + 40 t^3 u + 56 t^2 u^2 + 28 t u^3 + 5 u^4)
     + 4 t^2 u^2 (t + u)^5]\},
\nonumber\\
L&=& 18 v_c^2 Q^4 \{4 Q^6 t^2 [s^5 t - s^4 u^2 - 2 s^3 t^3 + s^2 
     t^2 u 
     (t + 3 u) + s t u (2 t^3 + t^2 u - t u^2 - u^3)
\nonumber\\ 
 & & {}- t^2 u^3 (2 t
     + u)] - Q^4 t [2 s^7 (t + u) + 4 s^6 (2 t^2 + u^2) 
     -s^5 (t^3 - t^2 u + 2 t u^2 - 8 u^3)
     -s^4 (21 t^4
\nonumber\\
 & & {}+ 13 t^3 u + 18 t^2 u^2
     - 2 t u^3 - 4 u^4)
     -2 s^3 (6 t^5 + 10 t^4 u + t^3 u^2 + 4 t^2 u^3 - 6 t u^4 - u^5)
     +2 s^2 t 
\nonumber\\
 & & {} \times (t^5 + 10 t^4 u + 11 t^3 u^2 + 9 t^2 u^3 + 6 t u^4 
     + 5 u^5)
     +s t^2 u (12 t^4 + 26 t^3 u + 2 t^2 u^2 + t u^3 + u^4)
\nonumber\\
 & & {}+t^3 u^2 (2 t^3 - 6 t^2 u - 15 t u^2 - 7 u^3)] 
     + 2 Q^2 [s^8 (t^2 - u^2) + s^7 (3 t^3 + t u^2 - 2 u^3)
     - s^6 (2 t^4
\nonumber\\
 & & {}+ 2 t^3 u + 2 t^2 u^2 - 
     t u^3 + 3 u^4)
     - s^5 (12 t^5 + 16 t^4 u + 16 t^3 u^2 - t^2 u^3 + 
     t u^4 + 2 u^5)
     - s^4 (10 t^6
\nonumber\\
 & & {}+ 28 t^5 u + 28 t^4 u^2 + 13 t^3 u^3 - 7 t^2 u^4 + 3 t u^5 
     + u^6)- s^3 t (t^6 + 9 t^5 u + 18 t^4 u^2 + t^3 u^3
\nonumber\\
 & & {}- 15 t^2 u^4 - 10 t u^5 + 
     2 u^6)
     + s^2 t^2 (t^6 + 7 t^5 u + 13 t^4 u^2 + 18 t^3 u^3 + 
     35 t^2 u^4 + 24 t u^5 + 6 u^6)
\nonumber\\
 & & {}+ s t^3 u (2 t^5 + 10 t^4 u + 16 t^3 u^2 + 
     20 t^2 u^3 + 19 t u^4 + 7 u^5)
     + t^6 u^2 (t + u)^2] 
     + s (s + u) [2 s^7 u (t
\nonumber\\
 & & {}+ u) + 2 s^6 u (3 t^2 + 3 t u + 2 u^2)
     + s^5 (5 t^4 + 15 t^3 u + 18 t^2 u^2 + 
     14 t u^3 + 6 u^4)
     + s^4 (15 t^5 
\nonumber\\
 & & {}+ 38 t^4 u + 53 t^3 u^2 + 
     40 t^2 u^3 + 22 t u^4 + 4 u^5)
     + s^3 (15 t^6 + 52 t^5 u + 88 t^4 u^2 + 81 t^3 u^3 + 47 t^2 u^4
\nonumber\\
 & & {}+ 19 t u^5 + 2 u^6)
     + s^2 t (5 t^6 + 32 t^5 u + 78 t^4 u^2 + 
     90 t^3 u^3 + 68 t^2 u^4 + 34 t u^5 + 9 u^6)
     + s t^2 u (7 t^5
\nonumber\\
 & & {}+ 31 t^4 u + 
     47 t^3 u^2 + 39 t^2 u^3 + 23 t u^4 + 7 u^5)
     + 2 t^5 u^2 (t + u)^2]\} 
     - 10 a_c^2 t \{
     4 Q^{10} t^4 [2 s^3 
\nonumber\\
 & & {}+ s^2 (5 t + 7 u) + s (4 t^2 + 11 t u + 7 u^2)
     + (t + u)^2 (t + 2 u)] 
     - 2 Q^8 t^2 [2 s^6 + 2 s^5 (4 t + 5 u) 
\nonumber\\
 & & {}+ 2 s^4 (9 t^2 + 11 t u + 9 u^2) + 
     s^3 (29 t^3 + 43 t^2 u + 15 t u^2 + 15 u^3) + 
     s^2 (27 t^4 + 73 t^3 u + 41 t^2 u^2
\nonumber\\
 & & {}+ t u^3 + 6 u^4) + 
     s (12 t^5 + 52 t^4 u + 73 t^3 u^2 + 33 t^2 u^3 + t u^4 + u^5) + 
     t (2 t^5 + 12 t^4 u + 27 t^3 u^2
\nonumber\\
 & & {}+ 29 t^2 u^3 + 13 t u^4 + u^5)]
     + Q^6 t [s^7 (11 t - 4 u) + 5 s^6 (9 t^2 + 8 t u - 4 u^2) + 
     s^5 (81 t^3 + 153 t^2 u
\nonumber\\
 & & {}+ 73 t u^2 - 36 u^3) + 
     s^4 (87 t^4 + 221 t^3 u + 226 t^2 u^2 + 99 t u^3 - 30 u^4) 
     + s^3 (56 t^5 + 198 t^4 u 
\nonumber\\
 & & {}+ 231 t^3 u^2 + 183 t^2 u^3 + 82 t u^4 - 
     12 u^5) + s^2 (14 t^6 + 106 t^5 u + 201 t^4 u^2 + 145 t^3 u^3 + 
     69 t^2 u^4
\nonumber\\
 & & {}+ 31 t u^5 - 2 u^6) - 2 s t (2 t^6 - 8 t^5 u - 
     53 t^4 u^2 - 82 t^3 u^3 - 45 t^2 u^4 - 8 t u^5 - 2 u^6)
     - 2 t^2 (t + u)^2
\nonumber\\
 & & {} \times (t^4 - 8 t^2 u^2 - 12 t u^3 - 3 u^4)]
     - Q^4 [s^9 (t + u) + s^8 (14 t^2 - t u + 6 u^2)
     + s^7 (48 t^3 + 35 t^2 u 
\nonumber\\
 & & {}- 18 t u^2 + 16 u^3)
     + s^6 (78 t^4 + 148 t^3 u + 17 t^2 u^2 - 54 t u^3 + 22 u^4)
     + s^5 (75 t^5 + 240 t^4 u + 214 t^3 u^2
\nonumber\\
 & & {}- 30 t^2 u^3 - 86 t u^4 + 
     16 u^5) 
     + s^4 (48 t^6 + 229 t^5 u + 360 t^4 u^2 + 216 t^3 u^3 - 
     43 t^2 u^4 - 74 t u^5 
\nonumber\\
 & & {}+ 6 u^6) + s^3 (20 t^7 + 152 t^6 u + 339 t^5 u^2 
     + 326 t^4 u^3
     + 134 t^3 u^4 - 17 t^2 u^5 - 31 t u^6 + u^7)
     + s^2 t (4 t^7
\nonumber\\
 & & {}+ 62 t^6 u + 
     210 t^5 u^2 + 275 t^4 u^3 + 150 t^3 u^4 + 28 t^2 u^5 - 4 t u^6 - 
     5 u^7)
     + 2 s t^2 u (t + u)^2 
     (5 t^4 + 21 t^3 u 
\nonumber\\
 & & {}+ 21 t^2 u^2 + 2 t u^3 - u^4)
     + 4 t^3 u^2 (t + u)^5]
     + Q^2 s [2 s^9 (t + u) + s^8 (12 t^2 + 12 t u + 13 u^2)
     + s^7 (28 t^3
\nonumber\\
 & & {}+ 41 t^2 u + 37 t u^2 + 37 u^3)
     + s^6 (32 t^4 + 73 t^3 u + 50 t^2 u^2 + 62 t u^3 + 60 u^4)
     + s^5 (18 t^5 + 59 t^4 u
\nonumber\\
 & & {}+ 29 t^3 u^2 - 
     17 t^2 u^3 + 48 t u^4 + 60 u^5)
     + s^4 (4 t^6 + 11 t^5 u - 41 t^4 u^2 - 146 t^3 u^3 - 133 t^2 u^4 + 
     2 t u^5
\nonumber\\
 & & {}+ 37 u^6)
     - s^3 u (10 t^6 + 84 t^5 u + 225 t^4 u^2 + 
     279 t^3 u^3 + 162 t^2 u^4 + 21 t u^5 - 13 u^6)
     - s^2 u (t + u)^2
\nonumber\\
 & & {} \times (4 t^5 + 40 t^4 u + 87 t^3 u^2 + 57 t^2 u^3 + 
     16 t u^4 - 2 u^5)
     - 2 s t u^2 (t + u)^3 
     (4 t^3 + 15 t^2 u + 9 t u^2 + u^3)
\nonumber\\
 & & {}- 4 t^2 u^3 (t + u)^5]
     - s^2 (s + u) (s + t + u)^2 
     [s^6 (t + u) + s^5 (3 t^2 + 4 t u + 4 u^2) 
     + s^4 (3 t^3 + 7 t^2 u
\nonumber\\
 & & {}+ 11 t u^2 + 
     7 u^3)
     + s^3 (t^4 + 6 t^3 u + 13 t^2 u^2 + 14 t u^3 + 7 u^4)     
     + s^2 u (2 t^4 + 8 t^3 u + 13 t^2 u^2 + 11 t u^3
\nonumber\\
 & & {}+ 4 u^4)     
     + s u^2 (t + u)^2 (2 t^2 + 2 t u + u^2)     
     + t u^3 (t + u)^3]\},
\nonumber\\
A&=&0.
\end{eqnarray}

\end{appendix}

\newpage
\begin{figure}[ht]
\begin{center}
\centerline{
\epsfig{figure=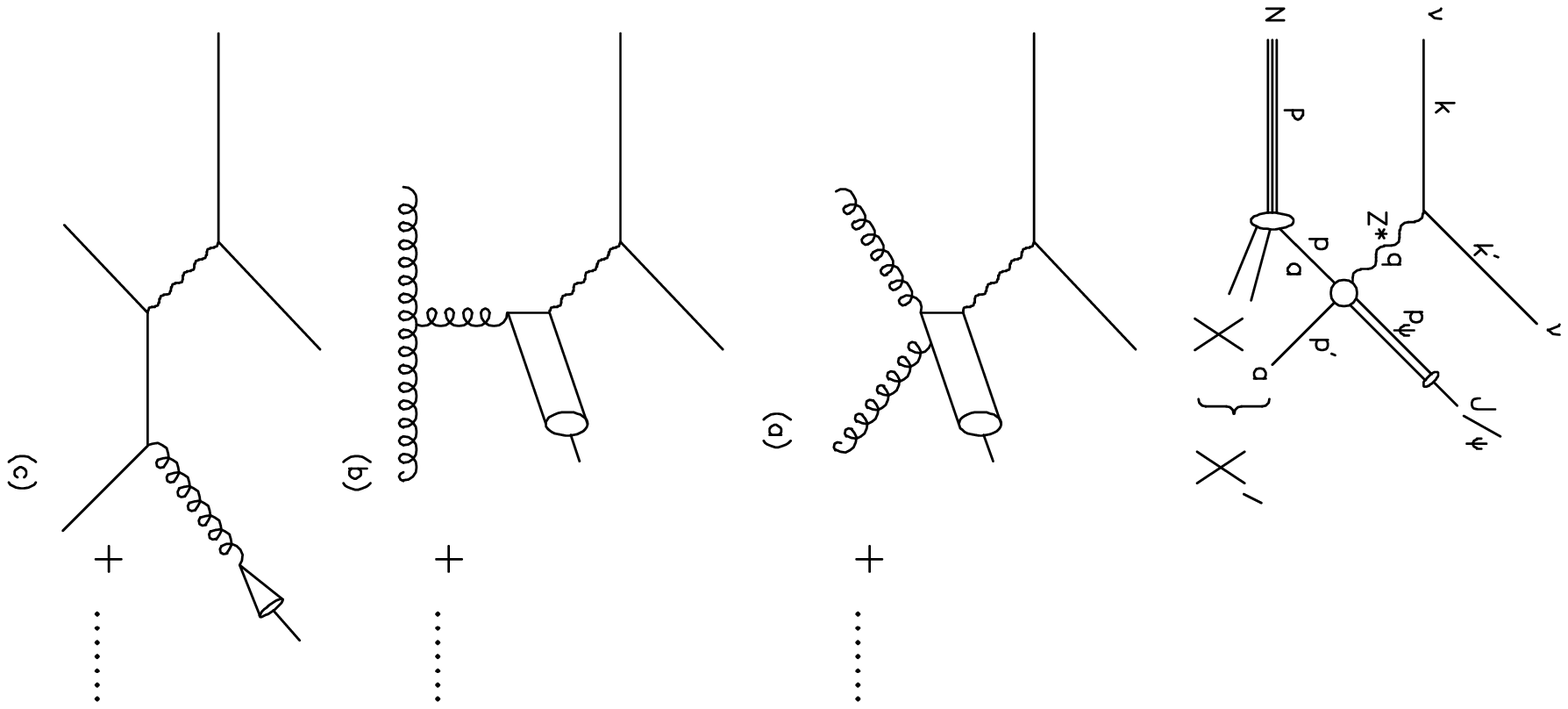,height=8cm,bbllx=39pt,bblly=505pt,%
bburx=415pt,bbury=729pt,angle=90,clip=}
}
\caption{Representative Feynman diagrams for the partonic subprocesses
$\nu+a\to\nu+c\overline{c}[n]+a$, where $a=q,\overline{q},g$ and
$n={}^3\!S_1^{(1)},{}^3\!P_J^{(1)},{}^1\!S_0^{(8)},{}^3\!S_1^{(8)},
{}^3\!P_J^{(8)}$.
There are six diagrams of the type shown in part (a), two ones of the type
shown in part (b), and two ones of the type shown in part (c).
There are two more diagrams that are obtained from the diagrams of the type
shown in part (b) by replacing the external gluon lines with quark ones.
The CS process only proceeds through the diagrams shown in part (a).}
\label{fig:fey}
\end{center}
\end{figure}

\newpage
\begin{figure}[ht]
\begin{center}
\centerline{
\epsfig{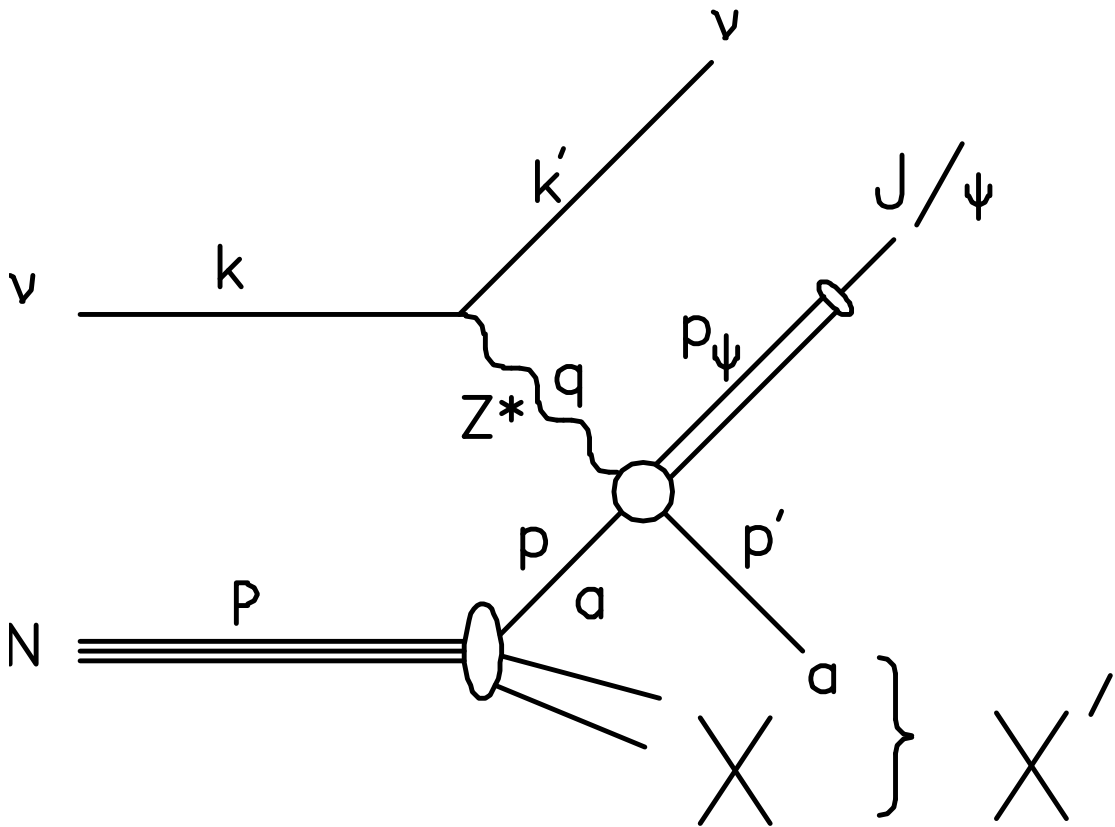}
}
\caption{Schematic representation of $\nu+N\to\nu+J/\psi+j+X$ explaining the
four-momentum assignments.}
\label{fig:kin}
\end{center}
\end{figure}

\newpage
\begin{figure}[ht]
  \begin{center}
    \setlength{\unitlength}{1cm}
    \begin{picture}(15,10)
      \setlength{\unitlength}{1cm}
      \put(0.5,0){\includegraphics[width=15cm,height=10cm]{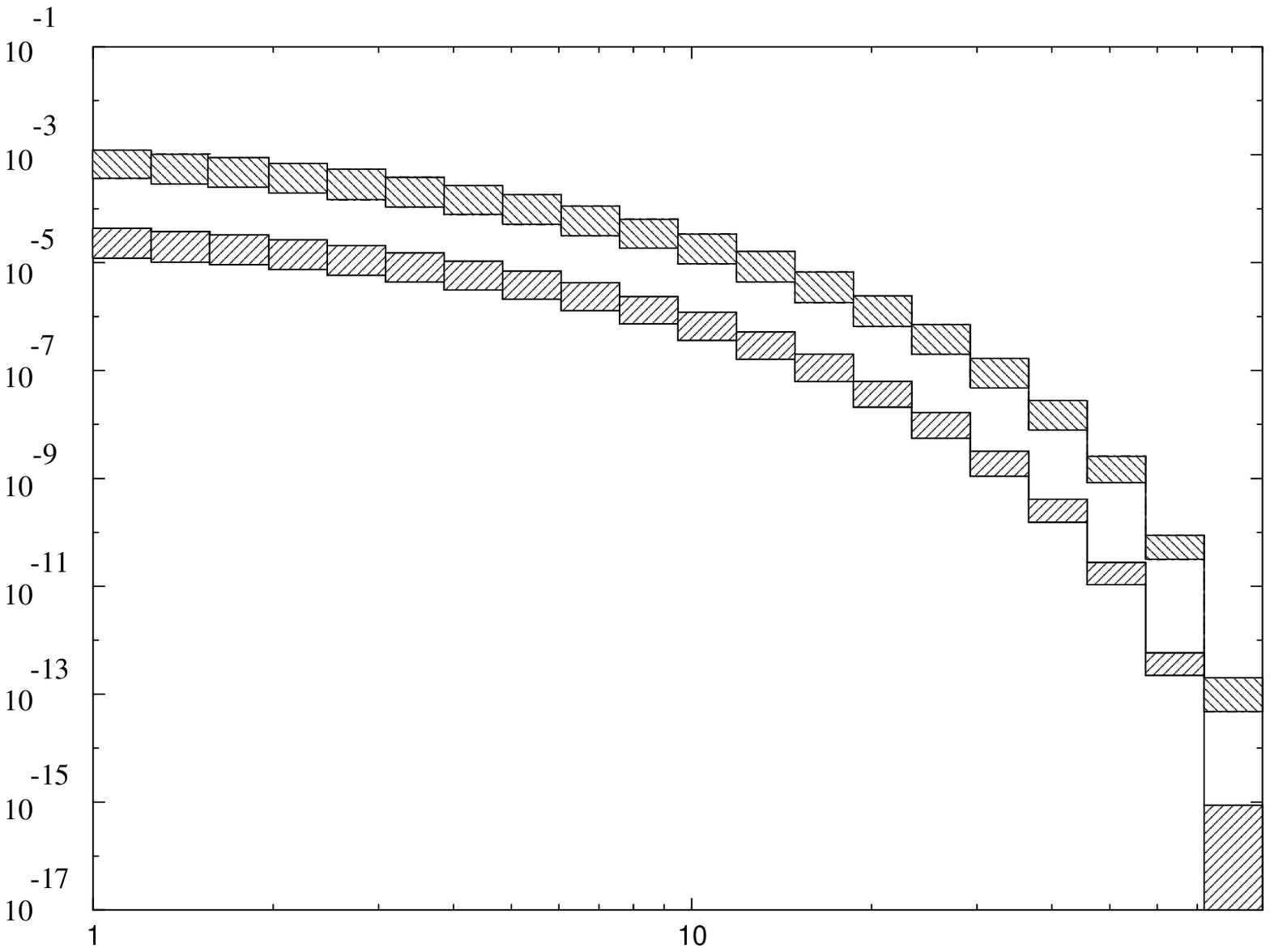}}
      \put(-1.5,5){$\displaystyle\frac{d\sigma}{dp_T^2}$}
      \put(-1.5,4){[fb/GeV${}^2$]}
      \put(7.5,-0.5){$p_T^2$ [GeV${}^2$]}
    \end{picture}
  \end{center}
\caption{LO NRQCD (upper histogram) and CSM (lower histogram) predictions for
the $p_T^2$ distribution of prompt $J/\psi$ inclusive production in
nondiffractive $\nu N$ NC DIS appropriate for the CHORUS and NOMAD
experiments.}
\label{fig:pT}
\end{figure}

\newpage
\begin{figure}[ht]
  \begin{center}
    \setlength{\unitlength}{1cm}
    \begin{picture}(15,10)
      \setlength{\unitlength}{1cm}
      \put(0.5,0){\includegraphics[width=15cm,height=10cm]{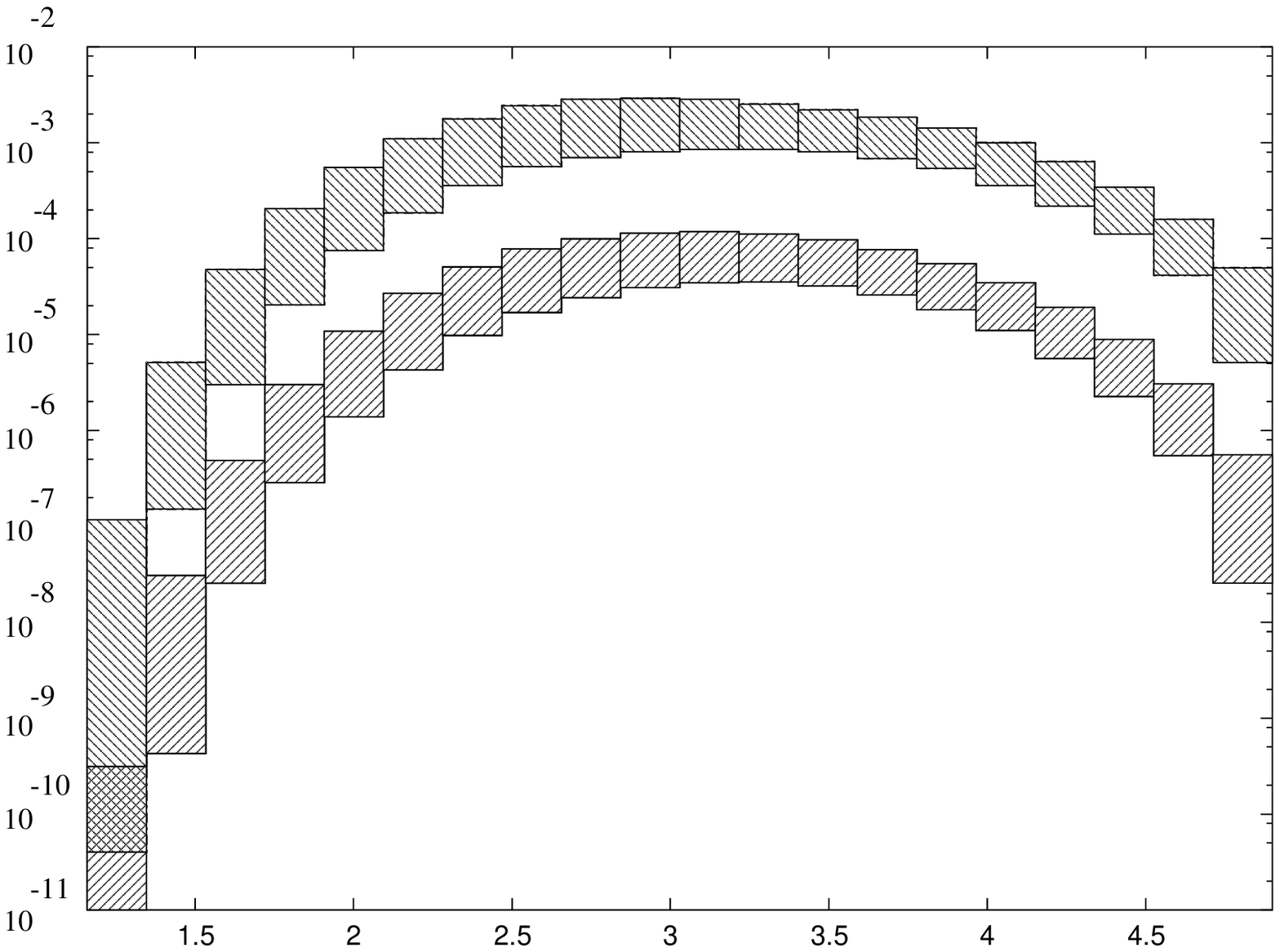}}
      \put(0,5){$\displaystyle\frac{d\sigma}{dy}$}
      \put(0,4){[fb]}
      \put(8.5,-0.5){$y$}
    \end{picture}
  \end{center}
\caption{LO NRQCD (upper histogram) and CSM (lower histogram) predictions for
the $y$ distribution of prompt $J/\psi$ inclusive production in nondiffractive
$\nu N$ NC DIS appropriate for the CHORUS and NOMAD experiments.}
\label{fig:y}
\end{figure}

\newpage
\begin{figure}[ht]
  \begin{center}
    \setlength{\unitlength}{1cm}
    \begin{picture}(15,10)
      \setlength{\unitlength}{1cm}
      \put(0.5,0){\includegraphics[width=15cm,height=10cm]{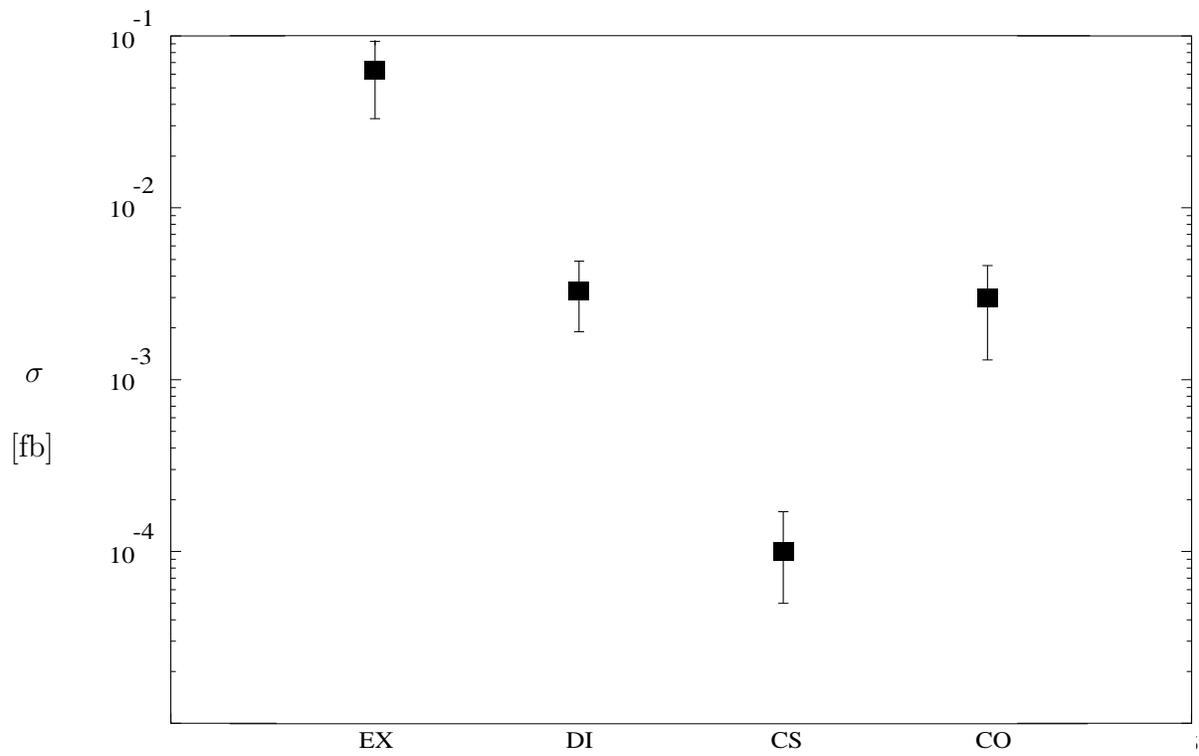}}
      \put(-0.1,5){$\sigma$}
      \put(-0.3,4){[fb]}
    \end{picture}
  \end{center}
\caption{The total cross section of $\nu+N\to\nu+J/\psi+X$ measured by CHORUS
\protect\cite{chorus} (EX) is compared with the predicted diffractive (DI),
CS, and CO contributions.}
\label{fig:exp}
\end{figure}

\end{document}